\documentclass[aps,pra,onecolumn]{revtex4}
\usepackage{graphicx}
\usepackage{dcolumn}
\usepackage{bm}
\usepackage{amsmath}
\usepackage{amssymb}
\usepackage{multirow}
\usepackage[misc]{ifsym}
\usepackage{soul}
\usepackage{xcolor}

\usepackage[bookmarks = false,
            colorlinks = true,
            linkcolor = blue,
            urlcolor  = black,
            citecolor = blue]{hyperref}
\usepackage{subeqnarray}
\allowdisplaybreaks

\newtheorem{definition}{Definition}[section]
\newtheorem{theorem}{Theorem}[section]
\newtheorem{lemma}{Lemma}[section]
\newtheorem{corollary}{Corollary}[section]

\newcommand{\bra}[1]{\langle#1|}
\newcommand{\ket}[1]{|#1\rangle}
\newcommand{\proj}[1]{|#1\rangle\!\langle#1|}
\newcommand{\braket}[2]{\langle#1|#2\rangle}
\newcommand{\ketbra}[2]{|#1\rangle\!\langle#2|}
\def\tr{{\rm tr}}
\def\det{{\rm det}}

\begin{document}
\title{An optimal discrimination of two mixed qubit states with a fixed rate of inconclusive results}
\author{Donghoon Ha and Younghun Kwon}
\email{yyhkwon@hanyang.ac.kr}
\affiliation{Department of Applied Physics, Hanyang University, Ansan, Republic of Korea}
\date{\today}

\begin{abstract}
In this paper we consider the optimal discrimination of two mixed qubit
states for a measurement that allows a fixed rate of inconclusive results. Our
strategy is to transform the problem of two qubit states into a minimum error discrimination for three qubit states by adding a specific quantum state $\rho_{0}$ and a prior probability $q_{0}$, which behaves as an inconclusive degree. First, we introduce the beginning and the end of practical interval of inconclusive result, $q_{0}^{(0)}$ and $q_{0}^{(1)}$, which are key ingredients in investigating our problem.
Then we obtain the analytic form of them. Next, we show that our problem can be classified into two cases $q_{0}=q_{0}^{(0)}$(or
$q_{0}=q_{0}^{(1)}$) and $q_{0}^{(0)}\!<\!q_{0}\!<\!q_{0}^{(1)}$. 
In fact, by maximum confidences of two qubit states and non-diagonal element of $\rho_{0}$, the our problem is completely understood. We provide an analytic solution of our problem when
$q_{0}=q_{0}^{(0)}$(or $q_{0}=q_{0}^{(1)}$). However, when $q_{0}^{(0)}\!<\!q_{0}\!<\!q_{0}^{(1)}$, we rather supply the numerical method to find the solution, because of the complex relation between inconclusive degree and corresponding failure probability. Finally we confirm our results using previously known examples.
\end{abstract}

\maketitle

\section{Introduction}
The information encoded in the quantum state by a sender can be
delivered to a receiver, who performs a measurement to extract this
information. A proper measurement strategy is required when the
receiver wants to obtain information from nonorthogonal quantum
states because those states cannot be perfectly discriminated\cite{ref:chef1,ref:barn1,ref:berg1,ref:bae1}. Measurement strategies can be classified by the constraints on conclusive or inconclusive results.
In quantum state discrimination, inconclusive results indicate that
the given quantum state cannot be definitely discriminated.
Minimum-error discrimination(MD)\cite{ref:hels1,ref:hole1,ref:yuen1,ref:ban1,ref:chou1,ref:herz1,ref:sams1,ref:deco1,ref:khia1,ref:jafa1,ref:bae2,ref:bae3,ref:ha1,ref:ha2} is able to minimize the average
error of conclusive results without inconclusive results.
Unambiguous discrimination(UD)\cite{ref:ivan1,ref:diek1,ref:pere1,ref:jaeg1,ref:rudo1,
ref:herz2,ref:pang1,ref:klei1,ref:sugi1,ref:berg2,ref:ha3} and
maximum-confidence discrimination(MC)\cite{ref:crok1} strategies permit
inconclusive results and minimize individual errors associated with
the conclusive results.\\
\indent In addition to these strategies, there is a scheme for
minimizing the average error of conclusive results while maintaining
a fixed rate of inconclusive results(FRIR)\cite{ref:chef2,ref:zhan1,ref:fiur1,ref:elda1,ref:herz3,ref:baga1,ref:naka1,ref:herz4}.
 FRIR is actually a generalization of other known
strategies. For example, when the fixed rate is zero, the FRIR is
equivalent to the MD. If the fixed rate is sufficiently large, the
FRIR becomes equivalent to MC(or UD). For the MD, the solution of two mixed
quantum states is explicitly known\cite{ref:hels1,ref:herz1}, however it is
not known for the FRIR. The solution of the FRIR of two qubit states
with identical maximal confidences exists\cite{ref:herz3} but that
of the general case does not.\\
\indent Recently, Bagan {\it et al.}\cite{ref:baga1} changed the FRIR of
$N$ quantum states into the MD of $N$ quantum states by modifying
prior probabilities and the quantum states. This approach can be
useful for obtaining a solution to symmetric states but it cannot be
used for arbitrary quantum states and prior probabilities because it
requires solving complicated equations. On the other hand, Nakahira
{\it et al.}\cite{ref:naka1} and Herzog\cite{ref:herz4} provided another
method to transform the FRIR into the MD. Their method does not
modify the given quantum states or the prior probabilities. Instead,
this method only adds an appropriate density operator $\rho_{0}$
with a suitable probability $q_{0}$(which we will call an
inconclusive degree) for a given quantum system. Then, based on the
FRIR of $N$ quantum states, one can form the MD of $N+1$ quantum
states by using a measurement operator that provides inconclusive
results. In order to transform the problem of optimal discrimination of $N$ quantum states
with a fixed rate of inconclusive results into that of minimum error discrimination of $N+1$ quantum
states, one must deal with special inconclusive degrees
$q_{0}^{(0)}$ and $q_{0}^{(1)}$. Even though they mentioned the relation between failure probability of
original problem and inconclusive degree of modified problem,
they could not find special inconclusive degrees
$q_{0}^{(0)}$ and $q_{0}^{(1)}$ in an analytic form, which appear
naturally in modified problem. Even more they could not solve even
the simplest FRIR problem for two qubit mixed states. Here special
inconclusive degrees $q_{0}^{(0)}$ and $q_{0}^{(1)}$ are the
beginning and the end of practical interval of inconclusive degree.
In fact $q_{0}^{(0)}$ and $q_{0}^{(1)}$ are the key to
solve FRIR of two qubit states.\\
\indent In fact, Nakahira
{\it et al.}\cite{ref:naka1} and Herzog\cite{ref:herz4} could not give a solution to the FRIR of two mixed qubit states. In this paper we provide a solution to the FRIR of two mixed qubit states. In Section
\ref{sec:FRIR} we derive the detailed relation between original FRIR
problem and modified FRIR problem, which is given by MD of three qubit states. Furthermore we introduce special inconclusive degrees
$q_{0}^{(0)}$ and $q_{0}^{(1)}$ and investigate their feature. In
Section \ref{sec:two} we divide FRIR problem of two qubit states
into two cases of $q_{0}=q_{0}^{(0)}$(or $q_{0}=q_{0}^{(1)}$) and
$q_{0}^{(0)}<q_{0}<q_{0}^{(1)}$, and specify that the problem can be solved by maximum confidences of
two qubit states and the non-diagonal element of $\rho_{0}$. Using
complementarity problem, we find the analytic form of $q_{0}^{(0)} $
and $q_{0}^{(1)}$, and provide the complete understanding of
modified FRIR problem in $q_{0}^{(0)}\leq q_{0}\leq q_{0}^{(1)}$.
That is, we provide an analytic solution of original FRIR problem in
case of $q_{0}=q_{0}^{(0)}$(or $q_{0}=q_{0}^{(1)}$). If
$q_{0}^{(0)}<q_{0}<q_{0}^{(1)}$, because of complex relation between
inconclusive degree and corresponding failure probability, we
provide the method to solve original problem numerically. Finally,
we confirm our results by providing the correct solutions to known
examples\cite{ref:herz3}. In Section \ref{sec:con} we summarize our
result.
\section{FRIR}\label{sec:FRIR}
We consider the quantum state ensemble
$\{q_{i},\rho_{i}\}_{i=1}^{N}$. This ensemble suggests that with the
prior probability $q_{i}$, one prepares the quantum state
corresponding to the density operator $\rho_{i}$ on a
$d$-dimensional complex Hilbert space $\mathcal{H}_{d}$. Without
loss of generality, we assume that the eigenvectors of
$\rho_{0}\equiv\sum_{i=1}^{N}q_{i}\rho_{i}$ (with nonzero
eigenvalues) span $\mathcal{H}_{d}$. The quantum state of the
system may be discriminated by the positive operator valued measure
(POVM) $\{M_{i}\}_{i=0}^{N}$. The POVM consists of $N+1$ positive
semidefinite Hermitian operators on $\mathcal{H}_{d}$ and satisfies
$\sum_{i=0}^{N}M_{i}=I_{d}$. $I_{d}$ is the identity operator on
$\mathcal{H}_{d}$. Here $M_{0}$ provides inconclusive results, while
$M_{i(\neq 0)}$ gives conclusive results. The probability that the
quantum state $\rho_{i}$ can be guessed to be $\rho_{j}$ is
$\tr[\rho_{i}M_{j}]$ by the Born rule. Therefore, the probability
for conclusive results $P_{\rm C}$ turns out to be
$\sum_{i=1}^{N}\tr[\rho_{0}M_{i}]$, and the probability for
inconclusive results $P_{\rm I}$ becomes $\tr[\rho_{0}M_{0}]$. The
probability of correctly guessing the quantum state and the error
probability are $P_{\rm cor}=\sum_{i=1}^{N}q_{i}\tr[\rho_{i}M_{i}]$
and $P_{\rm err}=P_{\rm C}-P_{\rm cor}$ respectively. We use $R_{\rm
cor(err)}$ to denote the probability of correctly(or incorrectly)
guessing when we succeed in guessing the quantum state. That is,
$R_{\rm cor(err)}=P_{\rm cor(err)}/P_{\rm C}$.
\subsection{Original FRIR problem}
Our discrimination strategy is to maximize(or minimize) $R_{\rm cor(err)}$ with fixed $P_{\rm I}=Q(0\leq Q<1)$.
Because $P_{\rm C}+P_{\rm I}=1$, this is equivalent to maximizing(or minimizing) $P_{\rm cor(err)}$ with fixed $P_{\rm I}=Q$,
which can be reformulated into the following optimization problem:
\begin{eqnarray}\label{eq:optp1}
\qquad\ {\rm max}&\quad& P_{\rm cor}=\sum_{i=1}^{N}q_{i}\tr[\rho_{i}M_{i}]\nonumber\\
{\rm subject\ to}&\quad& M_{i}\geq0\ \forall i,\ \sum_{i=0}^{N}M_{i}=I_{d},\
\tr[\rho_{0}M_{0}]=Q.
\end{eqnarray}
In this paper, we use the superscript
``opt'' to denote the optimized value or variable. For example, $P_{\rm
cor}^{\rm opt}(Q)$ and $R_{\rm cor}^{\rm opt}(Q)$ indicate the
maximum of $P_{\rm cor}$ and $R_{\rm cor}$ when $P_{\rm I}=Q$,
respectively.
\subsection{Modified FRIR problem}
Instead of simply attacking the problem as described above, we can
modify it as follows. Here, we introduce a positive number
$q_{0}$(called an inconclusive degree) which corresponds to the a
priori probability of $\rho_{0}$. Further, $M_{0}$ denotes the
measurement operator of guessing $\rho_{0}$ in the system:
\begin{eqnarray}\label{eq:optp2}
\qquad\ {\rm max}&\quad&\bar{P}_{\rm cor}\equiv\sum_{i=0}^{N}q_{i}\tr[\rho_{i}M_{i}]\nonumber\\
{\rm subject\ to}&\quad&M_{i}\geq0\ \forall i,\ \sum_{i=0}^{N}M_{i}=I_{d},\ q_{0}=q.
\end{eqnarray}
We use $\bar{P}_{\rm cor}^{\rm opt}(q)$ to
denote the maximum value of $\bar{P}_{\rm cor}$ when $q_{0}=q$.\\
\indent The following relation\cite{ref:naka1}
between ${P}_{\rm cor}^{\rm opt}(Q)$ and $\bar{P}_{\rm cor}^{\rm opt}(q)$
implies that when the MD of $\{q_{i},\rho_{i}\}_{i=0}^{N}$ can be completely analyzed, $R_{\rm cor}^{\rm
opt}(Q)$ can be found in the FRIR of $\{q_{i},\rho_{i}\}_{i=1}^{N}$.
\begin{lemma}\label{lem:rel}
If $P_{\rm I}=Q$ and $\bar{P}_{\rm cor}=\bar{P}_{\rm cor}^{\rm opt}(q)$ for some POVM,
$P_{\rm cor}^{\rm opt}(Q)=\bar{P}_{\rm cor}^{\rm opt}(q)-qQ$.
\end{lemma}
The proof is given in Appendix \ref{app:lem2}.\\ 
\indent Equation \eqref{eq:optp2} represents a convex optimization problem\cite{ref:boyd1}(or semidefinite program) to minimum-error discrimination problem for $\{q_{i},\rho_{i}\}_{i=0}^{N}$ with non-normalized priori probabilities. For investigating the analytic structure of POVM for an optimal solution of \eqref{eq:optp2}, we consider Karush-Kuhn-Tucker(KKT) optimality conditions, composed of constraints of primal and dual problem and complementary conditions, instead of necessary and sufficient conditions\cite{ref:hels1,ref:hole1,ref:yuen1}. 

\subsection{Optimality conditions of modified FRIR problem}
The optimization problem \eqref{eq:optp2} is equivalent to
MD of $\{q_{i},\rho_{i}\}_{i=0}^{N}$ with non-normalized priori probabilities.
Since the semidefinite programming of MD\cite{ref:elda2} hold regardless of the normalization condtion, we can apply the results into this modified FRIR problem.
First, the modified problem \eqref{eq:optp3} has the following Lagrange dual problem.
\begin{eqnarray}\label{eq:optp3}
{\rm min}\quad\tr[K]\quad{\rm subject\ to}\quad K=q_{i}\rho_{i}+r_{i}\tau_{i}\ \forall i ,\ q_{0}=q.
\end{eqnarray}
$K$ is a Hermitian operator on $\mathcal{H}_{d}$, which is a Lagrange multiplier of an equality constraint $\sum_{i=0}^{N}M_{i}=I_{d}$. $r_{i}\tau_{i}$ is a Lagrange multiplier of an inequality constraint $M_{i}\geq 0$, where $r_{i}$ and $\tau_{i}$ are a non-negative real number and a density operator on $\mathcal{H}_{d}$, respectively.
$r_{i}\tau_{i}$ is separated into $r_{i}$ and $\tau_{i}$, for  geometric understanding of qubit state discrimination.
Second, the optimized values of two
problems \eqref{eq:optp2},\eqref{eq:optp3} of $q_{0}=q$ coincide.
Finally, the complementary slackness condition
$r_{i}\tr[\tau_{i}M_{i}]=0(\forall i)$ is a necessary and sufficient
condition for optimizing the feasible variables in two
problems(primal and dual problems)\eqref{eq:optp2},\eqref{eq:optp3} of $q_{0}=q$. We summarize
the KKT optimality condition
for modified FRIR problem of $q_{0}=q$ as follows:
\begin{eqnarray}\label{eq:kkt}
&{\rm (i)}&\ {M}_{i}\geq0\ \forall i ,\ \sum_{i=0}^{N}{M}_{i}=I_{d},\nonumber\\
&{\rm (ii)}&\ q\rho_{0}+r_{0}\tau_{0}=q_{i}\rho_{i}+r_{i}\tau_{i}\ \forall i,\ \nonumber\\
&{\rm (iii)}&\ r_{i}\tr[\tau_{i}M_{i}]=0\ \forall i.
\end{eqnarray}
In order to express KKT optimality condition \eqref{eq:kkt} in a
form that we can deal with, we define the following variables.
\begin{eqnarray}\label{eq:demrt}
\bar{M}_{i}=\rho_{0}^{1/2}M_{i}\rho_{0}^{1/2},\
\bar{\rho}_{i}=\rho_{0}^{-1/2}q_{i}\rho_{i}\rho_{0}^{-1/2},\  \bar{\tau}_{i}=\rho_{0}^{-1/2}r_{i}\tau_{i}\rho_{0}^{-1/2}.
\end{eqnarray}
In terms of these newly defined variables, the KKT condition can be rewritten as:
\begin{eqnarray}\label{eq:optc1}
&{\rm (i)}&\ \bar{M}_{i}\geq0\ \forall i ,\ \sum_{i=0}^{N}\bar{M}_{i}=\rho_{0},\nonumber\\
&{\rm (ii)}&\ qI_{d}+\bar{\tau}_{0}=\bar{\rho}_{i}+\bar{\tau}_{i}\ \forall i,\nonumber\\
&{\rm (iii)}&\ \tr[\bar{\tau}_{i}\bar{M}_{i}]=0\ \forall i.~~
\end{eqnarray}
We denote $C_{i}$ and $\ket{\nu_{i}}$ as the largest eigenvalue of
$\bar{\rho}_{i}$ and the corresponding eigenvector, respectively.
$C_{i}$ physically represents the maximum achievable confidence of
$\rho_{i}$ in terms of MC\cite{ref:crok1}. Note that the product $r_{i}^{\rm
opt}\tau_{i}^{\rm opt}$ of $r_{i}^{\rm opt}$ and $\tau_{i}^{\rm
opt}$ satisfying optimality condition \eqref{eq:kkt} is unique, but
optimal POVM elements $M_{i}^{\rm opt}$ is not always
unique\cite{ref:bae3}. $\bar{\tau}_{i}^{\rm opt}$ fulfilling another
optimality condition \eqref{eq:optc1} is unique. However
$\bar{M}_{i}^{\rm opt}$ can be unique or non-unique. We will use the fact to find the analytic expression of optimal POVM element $M_{i}^{\rm opt}$
 or $\bar{M}_{i}^{\rm opt}$.\\
\indent When $d=2$, by introducing a real number $p_{i}$ and Bloch vectors ${\bm u}_{i}$, ${\bm v}_{i}$, and ${\bm w}_{i}$, we can express POVM elements $M_{i}$ and density operators $\rho_{i},\tau_{i}$ as:
\begin{eqnarray}\label{eq:gmea}
M_{i}=p_{i}(I_{2}+{\bm u}_{i}\cdot{\bm \sigma}),\
\rho_{i}=\frac{1}{2}(I_{2}+{\bm v}_{i}\cdot{\bm \sigma}),\
\tau_{i}=\frac{1}{2}(I_{2}+{\bm w}_{i}\cdot{\bm \sigma}).
\end{eqnarray}
Then, the KKT optimality condition \eqref{eq:kkt} can be described as:
\begin{eqnarray}\label{eq:optc2}
&{\rm (i)}&p_{i}\geq0\ \forall i,\ \sum_{i=0}^{N}p_{i}=1 ,\ \sum_{i=0}^{N}p_{i}{\bm u}_{i}=0, \nonumber\\
&{\rm (ii)}& q+r_{0}=q_{i}+r_{i},\ q{\bm v}_{0}+r_{0}{\bm w}_{0}=q_{i}{\bm v}_{i}+r_{i}{\bm w}_{i} \ \forall i, \nonumber\\
&{\rm (iii)}& p_{i}r_{i} (1+{\bm u}_{i}\cdot{\bm w}_{i})=0\ \forall i.
\end{eqnarray}
In Section \ref{sec:two} we investigate optimal variables of primal
problem \eqref{eq:optp2} and dual problem \eqref{eq:optp3}, using two
optimality conditions \eqref{eq:optc1} and \eqref{eq:optc2}. The
approach is called complementarity problem\cite{ref:bae2,ref:bae3}
in semidefinite programming.
\subsection{Special inconclusive degrees}
The fact that optimal measurement may not be unique in the MD leads us to introduce the following definition.
\begin{definition}
When $q$ is a positive number, we define $P_{\rm I}(q)$ as follows:
\begin{eqnarray}
P_{\rm I}(q)=\Bigg\{\tr[\rho_{0}M_{0}]&:& M_{i}\geq 0\ \forall i,\,\sum_{i=0}^{N}M_{i}=I_{d},\,q\,\tr[\rho_{0}M_{0}]+\sum_{i=1}^{N}q_{i}\tr[\rho_{i}M_{i}]=\bar{P}_{\rm cor}^{\rm opt}(q)\Bigg\}.
\end{eqnarray}
\end{definition}
The case of $Q\geq\!(=)a$ for any $Q\in P_{\rm I}(q)$ will be
denoted as $P_{\rm I}(q)\geq\!(=)a$, whereas that of $Q\geq\!(=)
Q'$ for any $Q\in P_{\rm I}(q)$ and $Q'\in P_{\rm I}(q')$ will be
written as $P_{\rm I}(q)\geq\!(=) P_{\rm
I}(q')$. Note that $0\leq P_{\rm I}(q) \leq 1$ for any $q$.\\
\indent The following lemma shows how $P_{\rm I}(q)$ behaves as $q$
increases.
\begin{lemma}\label{lem:con}
$P_{\rm I}(q)$ is a convex set for any $q$, and $P_{\rm I}(q)\leq P_{\rm I}(q')$  for any $q,q'$ with $q<q'$.
\end{lemma}
The proof is given in Appendix \ref{app:lem2}. Through Lemma \ref{lem:con}, $P_{\rm I}(q)$ is generally an interval. However, when optimal measurement of modified FRIR problem
$\{q_{i},\rho_{i}\}_{i=0}^{N}(q_{0}=q)$ is unique, $P_{\rm I}(q)$ becomes a point.\\
\indent Lemma \ref{lem:con} enables us to define the following special inconclusive degrees.
\begin{definition}[special inconclusive degrees]
We define $q_{0}^{(0)},q_{0}^{(1)}$ as follows:
\begin{eqnarray}
q_{0}^{(0)}&=&\max\{q>0:0\in P_{\rm I}(q)\},\nonumber\\
q_{0}^{(1)}&=&\min\{q>0:1\in P_{\rm I}(q)\}.
\end{eqnarray}
\end{definition}
This implies that a proper inconclusive degree $q$, which satisfies
$0<P_{\rm I}(q)<1$, exists in the region
$[q_{0}^{(0)},q_{0}^{(1)}]$. Therefore, $R_{\rm cor}^{\rm opt}(Q)$
in $0\leq Q<1$ can be found from $P_{\rm I}(q)$ and $\bar{P}_{\rm
cor}^{\rm opt}(q)$ in $q_{0}^{(0)}\leq q \leq q_{0}^{(1)}$. That is,
\begin{eqnarray}\label{eq:rbarp}
R_{\rm cor}^{\rm opt}(Q)=\frac{\bar{P}_{\rm cor}^{\rm opt}(q)-qQ}{1-Q}\quad\forall Q\in P_{\rm I}(q).
\end{eqnarray}
\indent The following lemma provides the lower bound of $q_0^{(0)}$ and
the upper bound of $q_{0}^{(1)}$.
\begin{lemma}\label{lem:eq}
$q_{0}^{(0)} \geq 1/N$ and $q_{0}^{(1)}\leq \max_{i}C_{i}$.
\end{lemma}
The proof is given in Appendix \ref{app:lem2}.
\section{Main Result:\ FRIR of two qubit mixed states}\label{sec:two}
In this section we analyze the FRIR of two qubit-mixed
states($d=N=2$), using the transformed KKT optimality condition which has two different forms.
The first KKT optimality condition \eqref{eq:optc1} is obtained by $\bar{M}_{i},\bar{\rho}_{i},\bar{\tau}_{i}$ of Eq. \eqref{eq:demrt}. The second one \eqref{eq:optc2} is expressed by Bloch vectors ${\bm u}_{i},{\bm v}_{i},{\bm w}_{i}$ defined in Eq. \eqref{eq:gmea}.  In certain situations, \eqref{eq:optc1} or \eqref{eq:optc2} is used. 
For two qubit-mixed states, $\bar{\rho}_{1}$,$\bar{\rho}_{2}$ are two positive semidefinite Hermitian operators on two-dimensional Hilbert space and they satisfy
$\bar{\rho_{1}}+\bar{\rho_{2}}=I_{2}$. $(1-C_{1})$ and $\ket{\nu_{1}}$($(1-C_{2})$ and $\ket{\nu_{2}}$) are the smallest eigenvalue and the corresponding eigenvector of $\bar{\rho}_{2}$($\bar{\rho}_{1}$), which implies $\braket{\nu_{1}}{\nu_{2}}=0$.
Therefore $\bar{\rho}_{1}$ and
$\bar{\rho}_{2}$ become:
\begin{eqnarray}\label{eq:barrho}
\bar{\rho}_{1}&=&C_{1} \proj{\nu_{1}}+(1-C_{2})\proj{\nu_{2}},\nonumber\\
\bar{\rho}_{2}&=&(1-C_{1}) \proj{\nu_{1}}+C_{2}\proj{\nu_{2}}.
\end{eqnarray}
We assume $C_1\leq C_{2}$, which does not spoil generality of the
problem. Here $e$ denotes the difference between $q_{1}$ and
$q_{2}$, and $l$ expresses the distance between two weighted Bloch
vectors $q_{1}{\bm v}_{1} $ and $q_{2}{\bm v}_{2}$.
\begin{eqnarray}
e=|q_{1}-q_{2}|,\ l=\|q_{1}{\bm v}_{1}-q_{2}{\bm v}_{2}\|_{2}.
\end{eqnarray}
\indent In this section, we divide our problem into two cases, by using two special inconclusive degrees $q_{0}^{(0)}$,$q_{0}^{(1)}$.
 In Subsection \ref{subsec:q01} and \ref{subsec:q00}, when fixed rate $Q$ of inconclusive results belongs to $P_{\rm I}(q_{0}^{(1)})$ or $P_{\rm I}(q_{0}^{(0)})$, we obtain what are optimal value $R_{\rm cor}^{\rm opt}$ and optimal measurement operators $M_{i}^{\rm opt}$(or $\bar{M}_{i}^{\rm opt}$). In Subsection \ref{subsec:q001}, when fixed rate lies in the other region($Q\in P_{\rm I}(q_{0}^{(0)})^{\sf C}\cap P_{\rm I}(q_{0}^{(1)})^{\sf C}$), we explain how complex optimal solution can be found. The result obtained from KKT optimality condition \eqref{eq:optc1} or \eqref{eq:optc2} is classified, according to the relation of two maximum confidences $C_{1}$,$C_{2}$ and that of $\rho_{11}$,$\rho_{12}$,$\rho_{22}$ of $\rho_{0}$.\\
\indent More specifically, in Subsection \ref{subsec:q01}, Corollary \ref{cor:q01}, which is the result of the section, is expressed in two cases, according to the equality between $C_{1}$ and $C_{2}$. Specially, the result of the case of $C_{1}=C_{2}$ is shown in three types, according the magnitude of three nonnegative numbers $\rho_{11},\rho_{22},|\rho_{12}|$.
In Subsection \ref{subsec:q00}, Theorem \ref{the:q00}, which is the final result of the section, is obtained in two cases,  by comparision between $\frac{1}{2}$ and $C_{1}$.
When $\frac{1}{2}<C_{1}$, the result is classified into two cases, by the existence of non-diagonal element $\rho_{12}$ of $\rho_{0}$. 
In Subsection \ref{subsec:q001}, the case of $\rho_{12}=0$ provides Theorem \ref{the:rho120} and that of $\rho_{12}\neq0$ gives Theorem \ref{the:q0q1m}. The former one is the result corresponding to the total range of fixed rate $Q$(that is, $0\leq Q\leq 1$). The latter one is the case of $Q\in P_{\rm I}(q_{0}^{(0)})^{\sf C}\cap P_{\rm I}(q_{0}^{(1)})^{\sf C}$.
\subsection{FRIR at $P_{\rm I}=Q$ for all $Q\in P_{\rm I}(q_{0}^{(1)})$}\label{subsec:q01}
In the following lemma modified FRIR problem to the case of
$q_{0}=C_{2}$ is completely analyzed.
\begin{lemma}\label{lem:q01}
$\bar{P}_{\rm cor}^{\rm opt}(C_{2})$ is $C_{2}$. When $C_{1}=C_{2}$,
$\bar{M}_{i}^{\rm opt}$ to $q_{0}=C_{2}$ is expressed as
\begin{eqnarray}
\bar{M}_{0}^{\rm opt}&=&\rho_{0}-\alpha\proj{\nu_{1}}-\beta\proj{\nu_{2}},\nonumber\\
\bar{M}_{1}^{\rm opt}&=&\alpha\proj{\nu_{1}},~~0\leq\alpha\leq\rho_{11},~0\leq\beta\leq\rho_{22},\nonumber\\
\bar{M}_{2}^{\rm opt}&=&\beta\proj{\nu_{2}},~~(\rho_{11}-\alpha)(\rho_{22}-\beta)\geq |\rho_{12}|^{2},
\end{eqnarray}
where
\begin{eqnarray}
\rho_{ij}=\bra{\nu_{i}}\rho_{0}\ket{\nu_{j}}.
\end{eqnarray}
Then $P_{\rm I}(C_{2})$ becomes
\begin{eqnarray}\label{eq:q01pi}
P_{\rm I}(C_{2})=
\left\{
\begin{array}{cll}
\mbox{$[Q_{1},1]$} &{\rm if}&\rho_{11}<|\rho_{12}|\leq\rho_{22},\\
\mbox{$[Q_{2},1]$} &{\rm if}&\rho_{22}<|\rho_{12}|\leq\rho_{11},\\
\mbox{$[2|\rho_{12}|,1]$} &{\rm if}&|\rho_{12}|\leq\rho_{11},\rho_{22},
\end{array}
\right.
\end{eqnarray}
where
\begin{eqnarray}
Q_{i}=\rho_{ii}+\frac{| \rho_{12}|^{2}}{\rho_{ii}}.
\end{eqnarray}
However when $C_{1}<C_{2}$, $P_{\rm I}(C_{2})$ becomes $[Q_{1},1]$,
and $\bar{M}_{i}^{\rm opt}$ for $q_{0}=C_{2}$ is expressed as
\begin{eqnarray}
\bar{M}_{0}^{\rm opt}&=&\rho_{0}-\beta\proj{\nu_{2}},\
\bar{M}_{1}^{\rm opt}=0,\nonumber\\
\bar{M}_{2}^{\rm opt}&=&\beta\proj{\nu_{2}},\
0\leq\beta\leq 1-Q_{1}.
\end{eqnarray}
\end{lemma}
The proof is given in Appendix \ref{app:lem3}. Note that three real numbers $Q_{1}$,\,$Q_{2}$, and $2|\rho_{12}|$ are less than 1. 
By Lemma \ref{lem:con}, $1\in P_{\rm I}(q)$ implies $q_{0}^{(1)}=q$ and $1\in P_{\rm I}(C_{2})$ derived by Lemma \ref{lem:q01} means $q_{0}^{(1)}=C_{2}$. Therefore, when $d=N=2$, inequality $q_{0}^{(1)}\leq\max_{i}C_{i}$ of Lemma \ref{lem:eq} becomes an equality.\\ 
\indent Lemma \ref{lem:q01} tells how the analytic solution of original FRIR problem is changed according to $Q\in P_{\rm I}(q_{0}^{(1)})$.
$\bar{P}_{\rm cor}^{\rm opt}(C_{2})$ is $C_{2}$ in any case and $R_{\rm cor}^{\rm opt}(Q)$ is $C_{2}$ for any $Q\in P_{\rm I}(C_{2})$, because of Eq. \eqref{eq:rbarp}. In case of $C_{1}=C_{2}$, modified FRIR measurement is represented by two variables $\alpha$ and $\beta$. However, in case of $C_{1}<C_{2}$, modified FRIR measurement is expressed only by $\beta$. This can be understood in terms of uniqueness of  FRIR measurement. In case of $C_{1}=C_{2}$, since $\tr[\bar{M}_{0}^{\rm opt}]=1-\alpha-\beta$, there may exist different $(\alpha,\beta)$ providing the same value of $\tr[\bar{M}_{0}^{\rm opt}]$, which implies that there are different forms of FRIR measurement in a fixed $P_{\rm I}$. However, in case of $C_{1}<C_{2}$, because of $\tr[\bar{M}_{0}^{\rm opt}]=1-\beta$, different $\beta$ provides different value of $\tr[\bar{M}_{0}^{\rm opt}]$. Therefore, according to $Q\in P_{\rm I}(q_{0}^{(1)})$, FRIR measurement uniquely exists. In other words, FRIR measurement of $C_{1}<C_{2}$ becomes FRIR measurement of $C_{1}=C_{2}$ with $\alpha=0$. Then, FRIR measurements of two cases can be represented by a variable $\epsilon=\rho_{11}-\alpha$. The following corollary summarizes the result.
\begin{corollary}[FRIR of $Q\in P_{\rm I}(q_{0}^{(1)})$]\label{cor:q01}
$q_{0}^{(1)}$ is $C_{2}$, and $P_{\rm I}(q_{0}^{(1)})$ can be
classified into
\begin{eqnarray} P_{\rm I}(q_{0}^{(1)})= \left\{
\begin{array}{cll}
\mbox{$[Q_{1},1]$} &{\rm if}&C_{1}<C_{2},\\
\mbox{$[Q_{1},1]$} &{\rm if}&C_{1}=C_{2},\ \rho_{11}<|\rho_{12}|\leq\rho_{22},\\
\mbox{$[Q_{2},1]$} &{\rm if}&C_{1}=C_{2},\ \rho_{22}<|\rho_{12}|\leq\rho_{11},\\
\mbox{$[2|\rho_{12}|,1]$} &{\rm if}&C_{1}=C_{2},\ |\rho_{12}|\leq\rho_{11},\rho_{22}.
\end{array}
\right.
\end{eqnarray}
$R_{\rm cor}^{\rm opt}(Q)$ is $C_{2}$, and $\bar{M}_{i}^{\rm opt}$
of $P_{\rm I}=Q$ can be expressed as
\begin{eqnarray}
\bar{M}_{0}^{\rm opt}&=&
\epsilon\proj{\nu_{1}}+\rho_{12}\ketbra{\nu_{1}}{\nu_{2}}+\rho_{21}\ketbra{\nu_{2}}{\nu_{1}}
+(Q-\epsilon)\proj{\nu_{2}},\nonumber\\
\bar{M}_{1}^{\rm opt}&=&(\rho_{11}-\epsilon)\proj{\nu_{1}},\
\bar{M}_{2}^{\rm opt}=(\rho_{22}-Q+\epsilon)\proj{\nu_{2}},
\end{eqnarray}
where
\begin{eqnarray}
\begin{array}{clll}
&\epsilon =\rho_{11}&{\rm if}&C_{1}<C_{2},\\
~\max\left\{Q-\rho_{22},\frac{Q}{2}-\sqrt{\frac{Q^2}{4}-|\rho_{12}|^{2}}\right\}\leq&\epsilon\leq\min\left\{\rho_{11},\frac{Q}{2}+\sqrt{\frac{Q^2}{4}-|\rho_{12}|^{2}}\right\}&{\rm if}&C_{1}=C_{2}.
\end{array}\nonumber
\end{eqnarray}
\end{corollary}
 The result implies that when $Q\in
P_{\rm I}(q_{0}^{(1)})$, if $C_{1}<C_{2}$, FRIR measurement to
$P_{\rm I}=Q$ is unique, but when $C_{1}=C_{2}$, it is not
unique. In the region of $P_{\rm I}(q_{0}^{(1)})$, though fixed rate $Q$ increases, $R_{\rm cor}^{\rm opt}(Q)$ is fixed as $C_{2}$ and the FRIR measurement has a unique form. The FRIR can be regarded as a MC. In the case of $C_{1}=C_{2}$, the FRIR, corresponding to the left-bound of $P_{\rm I}(q_{0}^{(1)})$, is equivalent to an optimal MC. In other words, when $C_{1}=C_{2}$, one of $Q_{1},Q_{2}$, and $2|\rho_{12}|$ becomes the minimum failure probability of MC, according to the relation of $\rho_{11},\rho_{22},\rho_{12}$ which are the component of $\rho_{0}$. However, since our strategy is to maximize average confidence at fixed failure probability, in case of $C_{1}<C_{2}$, the relation does not hold when $\rho_{11}<|\rho_{12}|\leq \rho_{22}$ is not satisfied.

\subsection{FRIR at $P_{\rm I}=Q$ for all $Q\in P_{\rm I}(q_{0}^{(0)})$}\label{subsec:q00}
When $q_{0}=q_{0}^{(0)}$, we classify modified FRIR problem into
three cases, using two maximum confidences and
the non-diagonal element $\rho_{12}$ of $\rho_{0}$. Then we analyze
the three cases completely. The first case is $C_{1}\leq \frac{1}{2}<C_{2}$,
and the second one $\frac{1}{2}<C_{1}\leq C_{2},\rho_{12}=0$. The third case
is $\frac{1}{2}<C_{1}\leq C_{2}$ and $\rho_{12}\neq 0$. The following lemma shows
the complete analysis to modified FRIR problem in $C_{1}\leq
\frac{1}{2}<C_{2}$ and $q_{0}=1-C_{1}$.
\begin{lemma}\label{lem:q001}
When $C_{1}\leq \frac{1}{2}<C_{2}$, $\bar{P}_{\rm cor}^{\rm opt}(1-C_{1})$
is $q_{2}$, and $P_{\rm I}(1-C_{1})$ is $[0,1-Q_{2}]$. If
$C_{1}<1/2$, $\bar{M}_{i}^{\rm opt}$ to $q_{0}=1-C_{1}$ becomes
\begin{eqnarray}
\bar{M}_{0}^{\rm opt}&=&\alpha\proj{\nu_{1}},\
\bar{M}_{1}^{\rm opt}=0,\nonumber\\
\bar{M}_{2}^{\rm opt}&=&\rho_{0}-\alpha\proj{\nu_{1}},\
0\leq\alpha\leq 1-Q_{2}.
\end{eqnarray}
However if $C_{1}=1/2$, they can be expressed as
\begin{alignat}{2}
&\bar{M}_{0}^{\rm opt}=\alpha\proj{\nu_{1}},\
\bar{M}_{1}^{\rm opt}=\beta\proj{\nu_{1}},\nonumber\\
&\bar{M}_{2}^{\rm opt}=\rho_{0}-(\alpha+\beta)\proj{\nu_{1}},\ \alpha,\beta\geq0,\ \alpha+\beta\leq 1-Q_{2}.
\end{alignat}
\end{lemma}
The proof is given in Appendix \ref{app:lem3}. Note that $1-Q_{2}$ is larger than zero. By Lemma \ref{lem:con}, $0\in P_{\rm I}(q)$ implies $q_{0}^{(0)}=q$ and by Lemma \ref{lem:q001}, $0\in P_{\rm I}(1-C_{1})$ means $q_{0}^{(0)}=1-C_{1}$. Therefore,
when $C_{1}\leq \frac{1}{2}<C_{2}$, if $C_{1}<\frac{1}{2}$, inequality $q_{0}^{(0)}\geq\frac{1}{2}$ of
Lemma \ref{lem:eq} becomes strictly an inequality, but if $C_{1}=\frac{1}{2}$, it becomes equality.\\ 
\indent Lemma \ref{lem:q001} shows how the analytic solution of original FRIR problem in the case of $C_{1}\leq \frac{1}{2}<C_{2}$ can be varied in terms of $Q\in P_{\rm I}(q_{0}^{(0)})$. In this case, $\bar{P}_{\rm cor}^{\rm opt}(1-C_{1})$ is $q_{2}$ and $R_{\rm cor}^{\rm opt}(Q)$ is $1-C_{1}+\frac{C_{1}-q_{1}}{1-Q}$ for any $Q\in P_{\rm I}(1-C_{1})$, because of Eq. \eqref{eq:rbarp}. However, the optimal measurements have different forms according to the case of $C_{1}<\frac{1}{2}$ or $C_{1}=\frac{1}{2}$. It is because in the case of $C_{1}<\frac{1}{2}$ the modified FRIR measurement is expressed only by a variable $\alpha$ but in the case of $C_{1}=\frac{1}{2}$ the modified FRIR measurement is given by $\alpha$ and $\beta$. In fact, FRIR measurement of the case of $C_{1}<\frac{1}{2}$ is equivalent to FRIR measurement of the case of $C_{1}=\frac{1}{2}$ with $\beta=0$. Therefore, by introducing a variable $\epsilon=\beta$, one can find the following corollary.
\begin{corollary}\label{cor:q002}
 When $C_{1}\leq \frac{1}{2}<C_{2}$, $q_{0}^{(0)}$ is $1-C_{1}$, and
$P_{\rm I}(q_{0}^{(0)})$ is $[0,1-Q_{2}]$. $\bar{M}_{i}^{\rm opt}$
for $P_{\rm I}=Q(\in P_{\rm I}(q_{0}^{(0)})$ can be expressed as
\begin{eqnarray}\label{eq:m1122}
\bar{M}_{0}^{\rm opt}&=&Q\proj{\nu_{1}},\ 
\bar{M}_{1}^{\rm opt}=\epsilon\proj{\nu_{1}},\nonumber\\
\bar{M}_{2}^{\rm opt}&=&(\rho_{11}-Q-\epsilon)\proj{\nu_{1}}+\rho_{12}\ketbra{\nu_{1}}{\nu_{2}}+\rho_{21}\ketbra{\nu_{2}}{\nu_{1}}+\rho_{22}\proj{\nu_{2}},~
\end{eqnarray}
where
\begin{eqnarray}
\begin{array}{llll}
&\epsilon =0&{\rm if}&C_{1}<\frac{1}{2}<C_{2},\\
0\leq&\epsilon\leq 1-Q_{2}-Q&{\rm if}&C_{1}=\frac{1}{2}<C_{2}.
\end{array}
\end{eqnarray}
\end{corollary}
The result tells that when $Q\in P_{\rm I}(q_{0}^{(0)})$,
if $C_{1}<1/2<C_{2}$, FRIR measurement for $P_{\rm I}=Q$ is unique,
but when $C_{1}=1/2<C_{2}$ it is not unique.\\
\indent The following lemma shows the solution to modified FRIR
problem of $\frac{1}{2}<C_{1}<C_{2}$,\,$\rho_{12}=0$,\,$q_{0}=C_{1}$.
\begin{lemma}\label{lem:q002}
 When $\frac{1}{2}<C_{1}\leq C_{2}$ and $\rho_{12}=0$, $\bar{P}_{\rm cor}^{\rm
opt}(C_{1})$ becomes $\rho_{11}C_{1}+\rho_{22}C_{2}$, and $P_{\rm
I}(C_{1})$ is $[0, \rho_{11}+\rho_{22}\delta_{C_{1},C_{2}}]$. If
$C_{1}<C_{2}$, $\bar{M}_{i}^{\rm opt}$ for $q_{0}=C_{1}$ is
expressed as
\begin{eqnarray}
\bar{M}_{0}^{\rm opt}&=&\alpha\proj{\nu_{1}},\nonumber\\ 
\bar{M}_{1}^{\rm opt}&=&(\rho_{11}-\alpha)\proj{\nu_{1}},\
0\leq\alpha\leq\rho_{11}\nonumber\\ 
\bar{M}_{2}^{\rm opt}&=&\rho_{22}\proj{\nu_{2}}.\label{eq:m31}
\end{eqnarray}
 However, if $C_{1}=C_{2}$, $\bar{M}_{i}^{\rm opt}$ for $q_{0}=C_{1}$ is given by
\begin{eqnarray}
\bar{M}_{0}^{\rm opt}&=&\alpha\proj{\nu_{1}}+\beta\proj{\nu_{2}},\nonumber\\
\bar{M}_{1}^{\rm opt}&=&(\rho_{11}-\alpha)\proj{\nu_{1}},\ 0\leq\alpha\leq\rho_{11},\nonumber\\ 
\bar{M}_{2}^{\rm opt}&=&(\rho_{22}-\beta)\proj{\nu_{2}},\ 0\leq\beta\leq\rho_{22}.\label{eq:m32}
\end{eqnarray}
\end{lemma}
The proof is given in Appendix \ref{app:lem3}. Note that $\rho_{11}$ is larger than zero. $0\in P_{\rm I}(C_{1})$ in Lemma \ref{lem:q002} includes $q_{0}^{(0)}=C_{1}$ by Lemma \ref{lem:con}. Therefore, when $\frac{1}{2}<C_{1}\leq C_{2}$, inequality $q_{0}^{(0)}\geq\frac{1}{2}$ in 
Lemma \ref{lem:eq} becomes strict.\\
\indent From Lemma \ref{lem:q002}, one can understand the behavior of analytic solution of original FRIR problem according to $Q\in P_{\rm I}(q_{0}^{(0)})$ when $\frac{1}{2}<C_{1}\leq C_{2}$ and $\rho_{12}=0$. In this case, $\bar{P}_{\rm cor}^{\rm opt}(C_{1})$ is $\rho_{11}C_{1}+\rho_{22}C_{2}$ and $R_{\rm cor}^{\rm opt}(Q)$ becomes $C_{1}+\frac{\rho_{22}(C_{2}-C_{1})}{1-Q}$ for any $Q\in P_{\rm I}(C_{1})$, because of Eq. \eqref{eq:rbarp}. When $C_{1}<C_{2}$, the modified FRIR measurement is expressed only by $\alpha$. However, when $C_{1}=C_{2}$, the modified FRIR measurement is given by $\alpha$ and $\beta$. In case of $C_{1}<C_{2}$, because of $\tr[\bar{M}_{0}^{\rm opt}]=\alpha$, the FRIR measurement is uniquely determined at a fixed $Q$. In case of $C_{1}=C_{2}$, because of $\tr[\bar{M}_{0}^{\rm opt}]=\alpha+\beta$, a fixed $Q$ cannot uniquely determine $\alpha$ and $\beta$. Therefore, it implies that FRIR measurement of $P_{\rm I}=Q$ may not be unique. The FRIR measurement of the case of $C_{1}<C_{2}$ may not be unique. FRIR measurement of the case of $C_{1}<C_{2}$ is the same as FRIR measurement of the case of $C_{1}=C_{2}$ with $\beta=0$.  Then, the following corollary can be obtained by introducing the variable $\epsilon=\beta$.
\begin{corollary}\label{cor:q003}
When $\frac{1}{2}<C_{1}\leq C_{2}$ and $\rho_{12}=0$, $q_{0}^{(0)}$ is
$C_{1}$, and $P_{\rm I}(q_{0}^{(0)})$ becomes
$[0,\rho_{11}+\rho_{22}\delta_{C_{1},C_{2}}]$. $\bar{M}_{i}^{\rm
opt}$ of $P_{\rm I}=Q(\in P_{\rm I}(q_{0}^{(0)}))$ is expressed as
\begin{eqnarray}\label{eq:m12120}
\bar{M}_{0}^{\rm opt}&=&(Q-\epsilon)\proj{\nu_{1}}+\epsilon\proj{\nu_{2}},\nonumber\\
\bar{M}_{1}^{\rm opt}&=&(\rho_{11}-Q+\epsilon)\proj{\nu_{1}},\nonumber\\
\bar{M}_{2}^{\rm opt}&=&(\rho_{22}-\epsilon)\proj{\nu_{2}},
\end{eqnarray}
where
\begin{eqnarray}
\begin{array}{clll}
 &\epsilon =0&{\rm if}&\frac{1}{2}<C_{1}< C_{2},\ \rho_{12}=0,\\
\max\{0,Q-\rho_{11}\}\leq&\epsilon\leq \min\{\rho_{22},Q\}&{\rm if}&\frac{1}{2}<C_{1}=C_{2},\ \rho_{12}=0.
\end{array}
\end{eqnarray}
\end{corollary}
 This result implies that when $Q\in P_{\rm I}(q_{0}^{(0)})$,
if $1/2<C_{1}<C_{2}$ and $\rho_{12}=0$, the FRIR measurement to $P_{\rm
I}=Q$ is unique. However, if $1/2<C_{1}=C_{2}$ and $\rho_{12}=0$, it
is not unique.\\ 
\indent From lemma \ref{lem:q004}, $q_{0}^{(0)}$ and
$P_{\rm I}(q_{0}^{(0)})$ can be found in modified FRIR problem to
$q_{0}=\chi$ when $1/2<C_{1}\leq C_{2}$ and $\rho_{12}\neq0$. Here
$\chi$ is as follows:
\begin{eqnarray}\label{eq:chi}
\chi=\frac{\chi_{1}+\chi_{2}-\sqrt{(\chi_{1}-\chi_{2})^{2}+4|\gamma_{12}|^{2}}}{2},
\end{eqnarray}
where
\begin{eqnarray}
\gamma_{ij}=\frac{l^{2}-e^{2}}{4l}\bra{\nu_{i}}\rho_{0}^{-1}\ket{\nu_{j}},\
\chi_{i}=\frac{1}{2}+\gamma_{ii}+\frac{(2q_{i}-1)(2C_{i}-1)}{2l}.
\end{eqnarray}
\begin{lemma}\label{lem:q004}
 When $\frac{1}{2}<C_{1}\leq C_{2}$ and $\rho_{12}\neq0$, we find
$q_{0}^{(0)}=\chi$, $\bar{P}_{\rm cor}^{\rm opt}(\chi)=\frac{1+l}{2}$, and
$P_{\rm I}(\chi)=0$. Then  $M_{i}^{\rm opt}$ to $P_{\rm I}=0$ is
expressed as
\begin{eqnarray}\label{eq:m1212}
M_{0}^{\rm opt}&=&0,\nonumber\\
M_{1}^{\rm opt}&=&\frac{1}{2}\left[I_{2}+\frac{(q_{1}{\bm v}_{1}-q_{2}{\bm v}_{2})\cdot{\bm\sigma}}{\|q_{1}{\bm v}_{1}-q_{2}{\bm v}_{2}\|_{2}}\right],\nonumber\\
M_{2}^{\rm opt}&=&\frac{1}{2}\left[I_{2}-\frac{(q_{1}{\bm v}_{1}-q_{2}{\bm v}_{2})\cdot{\bm\sigma}}{\|q_{1}{\bm v}_{1}-q_{2}{\bm v}_{2}\|_{2}}\right].\label{eq:mopt1}
\end{eqnarray}
\end{lemma}
The proof is given in Appendix \ref{app:lem3}. In Lemma \ref{lem:q004}, optimal POVM of $\{q_{i},\rho_{i}\}_{i=0}^{N}(q_{0}=\chi)$ is unique and we consider
$P_{\rm I}(\chi)$ not as a set $\{0\}$ but as a value $0$.\\
\indent The following theorem summarizes the previous results.
\begin{theorem}[FRIR of $Q\in P_{\rm I}(q_{0}^{(0)})$]\label{the:q00}
$q_{0}^{(0)}$ and $P_{\rm I}(q_{0}^{(0)})$ can be classified as
follows:
\begin{equation}
\begin{array}{llll}
q_{0}^{(0)}=1-C_{1},&P_{\rm I}(q_{0}^{(0)})=\mbox{$[0,1-Q_{2}]$} &{\rm if}&C_{1}\leq \frac{1}{2}<C_{2},\\
q_{0}^{(0)}=C_{1},&P_{\rm I}(q_{0}^{(0)})=\mbox{$[0,\rho_{11}+\rho_{22}\delta_{C_{1},C_{2}}]$} &{\rm if}&\frac{1}{2}<C_{1}\leq C_{2},\ \rho_{12}=0,\\
q_{0}^{(0)}=\chi,&P_{\rm I}(q_{0}^{(0)})=0 &{\rm if}&\frac{1}{2}<C_{1}\leq C_{2},\ \rho_{12}\neq 0.
\end{array}
\end{equation}
$R_{\rm cor}^{\rm opt}(Q)$ becomes
\begin{eqnarray}
R_{\rm cor}^{\rm opt}(Q)=
\left\{
\begin{array}{cll}
1-C_{1}+\frac{C_{1}-q_{1}}{1-Q} &{\rm if}&C_{1}\leq\frac{1}{2}<C_{2},\\
C_{1}+\frac{\rho_{22}(C_{2}-C_{1})}{1-Q} &{\rm if}&\frac{1}{2}<C_{1}\leq C_{2},\ \rho_{12}=0,\\
\frac{1+l}{2} &{\rm if}&\frac{1}{2}<C_{1}\leq C_{2},\ \rho_{12}\neq 0.
\end{array}
\right.\label{eq:relopt}
\end{eqnarray}
FRIR measurement of $P_{\rm I}=Q$ becomes, if $C_{1}\leq\frac{1}{2}
<C_{2}$, \eqref{eq:m1122}, and if $\frac{1}{2}<C_{1}\leq C_{2}$ and
$\rho_{12}=0$, is \eqref{eq:m12120}, and if $\frac{1}{2}<C_{1}\leq
C_{2}$ and $\rho_{12}\neq 0$, becomes \eqref{eq:m1212}.
\end{theorem}
When $Q$, corresponding to $P_{\rm I}=Q$, exists in $P_{\rm I}(q_{0}^{(0)})$, $R_{\rm cor}^{\rm opt}(Q)$ and the FRIR measurement have different forms in certain situations. In the case of $C_{1}\leq \frac{1}{2}<C_{2}$, $P_{\rm I}(q_{0}^{(0)})$ is neither $\{0\}$ nor $[0,1]$, because of $0<Q_{2}<1$. $P_{\rm I}$, corresponding to $q_{0}^{(0)}$ or $q_{0}^{(1)}$, exists not as a point but as an separate interval. It is not true in the case of $\frac{1}{2}<C_{1}\leq C_{2}$. If $\rho_{12}\neq 0$, $P_{\rm I}(q_{0}^{(0)})=\{0\}$. If $\rho_{12}=0$ and $C_{1}=C_{2}$, $P_{\rm I}(q_{0}^{(0)})=[0,1]$.  This implies that $q_{0}^{(0)}=q_{0}^{(1)}$. Then, the left-bound of $P_{\rm I}(q_{0}^{(1)})$ becomes 0. Since FRIR of $P_{\rm I}=0$ is MD, MD is an optimal MC when $\rho_{12}=0$ and $C_{1}=C_{2}$.
\subsection{FRIR at $P_{\rm I}=Q$ for all $Q\in P_{\rm I}(q_{0}^{(0)})^{\sf C}\cap P_{\rm I}(q_{0}^{(1)})^{\sf C}$}\label{subsec:q001}
\indent In the previous section we considered the case that the
failure probability $P_{\rm I}$ is fixed as $Q\in P_{\rm
I}(q_{0}^{(0)})\cap P_{\rm I}(q_{0}^{(1)})$. In this section, to
investigate FRIR in the other region, we classify modified FRIR
problem of $q_{0}^{(0)}<q_{0}<q_{0}^{(1)}$  into two cases. The
first case is $C_{1}\leq\frac{1}{2}<C_{2}$ and $\rho_{12}\neq0$, and the
second one is $\frac{1}{2}<C_{1}\leq C_{2}$ and $\rho_{12}\neq0$. Note that
$\rho_{12}=0$ is not included. It is because, from corollary
\ref{cor:q01} and theorem \ref{the:q00}, when $C_{1}=C_{2}$ and
$\rho_{12}=0$, we find $q_{0}^{(0)}=q_{0}^{(1)}$. When �
$C_{1}<C_{2}$ and $\rho_{12}=0$, we have
$q_{0}^{(0)}<q_{0}^{(1)}$. However, $P_{\rm
I}(q_{0}^{(0)})=[0,\rho_{11}]$ and $P_{\rm
I}(q_{0}^{(1)})=[\rho_{11},1]$ implies $P_{\rm
I}(q)=\rho_{11}(q_{0}^{(0)}<q<q_{0}^{(1)})$, which includes $R_{\rm
cor}^{\rm opt}(\rho_{11})=C_{2}$. In addition, the following lemma
tells that FRIR measurement to $P_{\rm I}=\rho_{11}$ is unique.
\begin{lemma}\label{lem:uni}
When $C_{1}<C_{2}$ and $\rho_{12}=0$, if
$q_{0}^{(0)}<q<q_{0}^{(1)}$, $\bar{M}_{i}^{\rm opt}$ at $q_{0}=q$
can be expressed as
\begin{eqnarray}\label{eq:m41}
\bar{M}_{0}^{\rm opt}=\rho_{11}\proj{\nu_{1}},\
\bar{M}_{1}^{\rm opt}=0,\
\bar{M}_{2}^{\rm opt}=\rho_{22}\proj{\nu_{2}}.
\end{eqnarray}
\end{lemma}
The proof is given in Appendix \ref{app:lem3}. The following theorem summarizes FRIR to $\rho_{12}=0$.
\begin{theorem}[FRIR of $\rho_{12}=0$]\label{the:rho120}
If $C_{1}=C_{2}$, $R_{\rm cor}^{\rm opt}(Q)$ becomes $C_{2}$ at any
$Q$, and $\bar{M}_{i}^{\rm opt}$ to $P_{\rm I}=Q$ can be representes
as
\begin{eqnarray}
\bar{M}_{0}^{\rm opt}&=&\epsilon_{1}\proj{\nu_{1}}+(Q-\epsilon_{1})\proj{\nu_{2}},\nonumber\\
\bar{M}_{1}^{\rm opt}&=&(\rho_{11}-\epsilon_{1})\proj{\nu_{1}},\nonumber\\ 
\bar{M}_{2}^{\rm opt}&=&(\rho_{22}-Q+\epsilon_{1})\proj{\nu_{2}},
\end{eqnarray}
where 
\begin{equation}
\max\{0,Q-\rho_{22}\}\leq\epsilon_{1}\leq\min\{\rho_{11},Q\}. 
\end{equation}
When $C_{1}<C_{2}$, if $0\leq Q\leq\rho_{11}$, $R_{\rm
cor}^{\rm opt}(Q)$ becomes
\begin{eqnarray}
R_{\rm cor}^{\rm opt}(Q)=
\left\{
\begin{array}{cll}
1-C_{1}+\frac{\rho_{22}(C_{1}+C_{2}-1)}{1-Q} &{\rm if}&C_{1}\leq\frac{1}{2},\\
C_{1}+\frac{\rho_{22}(C_{2}-C_{1})}{1-Q} &{\rm if}&\frac{1}{2}<C_{1}.
\end{array}
\right.
\end{eqnarray}
Then $\bar{M}_{i}^{\rm opt}$ for $P_{\rm I}=Q$ is expressed as
\begin{eqnarray}
\bar{M}_{0}^{\rm opt}&=&Q\proj{\nu_{1}},\ 
\bar{M}_{1}^{\rm opt}=\epsilon_{2}\proj{\nu_{1}},\nonumber\\ 
\bar{M}_{2}^{\rm opt}&=&(\rho_{11}-Q-\epsilon_{2})\proj{\nu_{1}}+\rho_{22}\proj{\nu_{2}},
\end{eqnarray}
where
\begin{eqnarray}
\begin{array}{llll}
&\epsilon_{2} =0&{\rm if}&C_{1}<\frac{1}{2}<C_{2},\\
0\leq&\epsilon_{2}\leq \rho_{11}-Q&{\rm if}&C_{1}=\frac{1}{2}<C_{2},\\
&\epsilon_{2}=\rho_{11}-Q&{\rm if}&\frac{1}{2}<C_{1}<C_{2}.
\end{array}
\end{eqnarray}
 If $\rho_{11}\leq Q<1$, $R_{\rm cor}^{\rm opt}(Q)$ is always
$C_{2}$, and $\bar{M}_{i}^{\rm opt}$ to $P_{\rm I}=Q$ becomes
\begin{eqnarray}
\bar{M}_{0}^{\rm opt}&=&\rho_{11}\proj{\nu_{1}}+(Q-\rho_{11})\proj{\nu_{2}},\nonumber\\
\bar{M}_{1}^{\rm opt}&=&0,~~ \bar{M}_{2}^{\rm opt}=(1-Q)\proj{\nu_{2}}.
\end{eqnarray}
\end{theorem}
When $\rho_{12}=0$, if $C_{1}=C_{2}$, MD becomes an optimal MC. When $C_{1}<C_{2}$, the right-bound of $P_{\rm I}(q_{0}^{(0)})$ is the same as the left-bound of $P_{\rm I}(q_{0}^{(1)})$, which happens only when $\rho_{12}=0$.

\begin{figure}[!t]
\centerline{\includegraphics[scale=0.50]{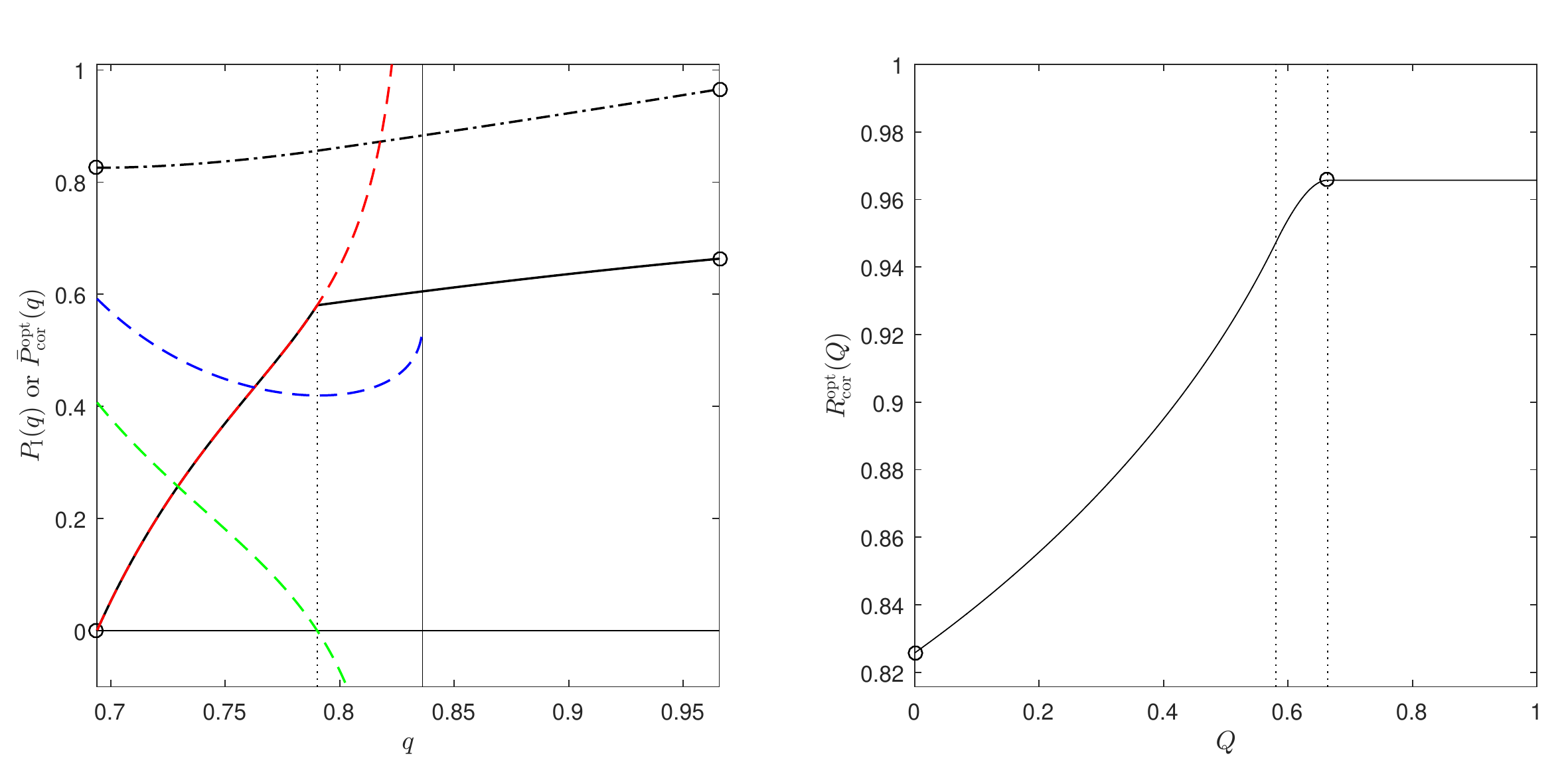}}\caption{The behavior
of $P_{\rm I}(q)$, $\bar{P}_{\rm cor}^{\rm opt}(q)$, and $R_{\rm
cor}^{\rm opt}(Q)$ at ${\bm v}_{1}=(-0.6,-0.2,-0.7)$ and ${\bm
v}_{2}=(-0.6,-0.1,0.6)$ with $q_{1}=0.4$ and $q_{2}=0.6$. We have
$|\rho_{12}|=0.3075$,\,$C_{1}=0.8361$,\,$C_{2}=0.9657$,
$q_{0}^{(0)}=\chi=0.6940$, and $q_{0}^{(1)}=C_{2}$. The left figure
shows $P_{\rm I}(q)$(solid line) and $\bar{P}_{\rm cor}^{\rm
opt}(q)$(dashed-dotted line) in $q_{0}^{(0)}<q<q_{0}^{(1)}$.
$\eta_{0}$(red dashed line) and $\eta_{2}$(blue dashed line) are
always positive in $q_{0}^{(0)}<q\leq C_{1}$. However,
$\eta_{1}$(green dashed line) is equal to or less than 0 in the
region $0.7902\leq q\leq C_{1}$. In $\chi< q_{0}<0.7902$ we get
$M_{i}^{\rm opt}\neq 0(\forall i)$, and in $0.7902\leq q_{0}<C_{2}$
we have $M_{0}^{\rm opt}\neq 0$, $M_{1}^{\rm opt}=0$, and
$M_{2}^{\rm opt}\neq0$. The right figure displays the behavior of
$R_{\rm cor}^{\rm opt}(Q)$, obtained from $P_{\rm I}(q)$ and
$\bar{P}_{\rm cor}^{\rm opt}(q)$ of the left figure. In $0<Q<0.5805$
we have $M_{i}^{\rm opt}\neq 0(\forall i)$, and in $0.5805\leq
Q<Q_{1}=0.6635$ we get $M_{0}^{\rm opt}\neq 0$, $M_{1}^{\rm opt}=0$,
and $M_{2}^{\rm opt}\neq0$.} \label{fig:ex}
\end{figure}

\indent The following theorem describes modified FRIR to
$\rho_{12}\neq0$ in $q_{0}\in(q_{0}^{(0)},q_{0}^{(1)})$.
\begin{theorem}\label{the:q0q1m}
When $\rho_{12}\neq 0$ and $q_{0}^{(0)}<q<q_{0}^{(1)}$, the optimal POVM
to $q_{0}=q$ is unique. Then at least one of $M_{1}^{\rm opt}$ and $M_{2}^{\rm opt}$ is nonzero. In
the case of $M_{x}^{\rm opt}\neq0$,\,$M_{y}^{\rm
opt}=0(\{x,y\}=\{1,2\})$, the index $x$ turns out to be the index
$i$ in $\max_{i\in{1,2}}[q_{i}+\|q{\bm v}_{0}-q_{i}{\bm
v}_{i}\|_{2}]$. In this case $\bar{P}_{\rm cor}^{\rm opt}(q)$ and
$P_{\rm I}(q)$ are given as
\begin{eqnarray}\label{eq:pcpi1}
\bar{P}_{\rm cor}^{\rm opt}(q)&=&
\frac{1}{2}\left(q+q_{x}+\|q{\bm v}_{0}-q_{x}{\bm v}_{x}\|_{2}\right),\nonumber\\
P_{\rm I}(q)&=&\frac{1}{2}\left[1+\frac{(q{\bm v}_{0}-q_{x}{\bm v}_{x})\cdot{\bm v}_{0}}{\|q{\bm v}_{0}-q_{x}{\bm v}_{x}\|_{2}}\right].~~~
\end{eqnarray}
The optimal POVM elements is represented as
\begin{eqnarray}\label{eq:optm1}
M_{0}^{\rm opt}&=&\frac{1}{2}\left[I_{2}+\frac{(q{\bm v}_{0}-q_{x}{\bm v}_{x})\cdot{\bm \sigma}}{\|q{\bm v}_{0}-q_{x}{\bm v}_{x}\|_{2}}\right],\nonumber\\
M_{x}^{\rm opt}&=&\frac{1}{2}\left[I_{2}+\frac{(q_{x}{\bm v}_{x}-q{\bm v}_{0})\cdot{\bm \sigma}}{\|q{\bm v}_{0}-q_{x}{\bm v}_{x}\|_{2}}\right],\,M_{y}^{\rm opt}=0.
\end{eqnarray}
If $M_{i}^{\rm opt}\neq 0(\forall i)$, $\bar{P}_{\rm cor}^{\rm
opt}(q)$ and ${P}_{\rm I}(q)$ become
\begin{equation}\label{eq:pcpi2}
\begin{array}{rll}
\bar{P}_{\rm cor}^{\rm opt}(q)&=&q+\rho_{11}\lambda_{1}+\rho_{22}\lambda_{2}
-2|\rho_{12}|\sqrt{\lambda_{1}\lambda_{2}},\\
P_{\rm I}(q)&=&
\Big[ \frac{(2C_{1}-1)(2C_{2}-1)-(2q-1)^{2}}{2(2q-1)^{2}(C_{1}+C_{2}-1)} \Big]\!\cdot\!\Big[\scalebox{0.75}{$ 1-2\rho_{11}C_{1} -2\rho_{22}C_{2}$}\\
&&+\frac{|\rho_{12}|(2C_{1}-1)(2C_{2}-1)[(C_{1}-q)(C_{2}-q)+(q-1+C_{1})(q-1+C_{2})]}{\sqrt{(2C_{1}-1)(2C_{2}-1)(C_{1}-q)(C_{2}-q)(q-1+C_{1})(q-1+C_{2})}}\Big],
\end{array}
\end{equation}
where
\begin{eqnarray}\label{eq:lam12}
\lambda_{1}&=&\frac{(2C_{2}-1)(C_{1}-q)(q-1+C_{1})}{(2q-1)(C_{1}+C_{2}-1)},\nonumber\\
\lambda_{2}&=&\frac{(2C_{1}-1)(C_{2}-q)(q-1+C_{2})}{(2q-1)(C_{1}+C_{2}-1)}.
\end{eqnarray}
Then $\{\bar{M}_{i}^{\rm opt}\}_{i=0}^{2}$ is expressed as
\begin{equation}\label{eq:optm2}
\begin{array}{rll}
\bar{M}_{0}^{\rm
opt}&=&\frac{\eta_{0}}{\lambda_{1}+\lambda_{2}}\Big[\scalebox{0.75}{$
\lambda_{2}\proj{\nu_{1}}+$}\frac{\rho_{12}\sqrt{\lambda_{1}\lambda_{2}}}{|\rho_{12}|}\scalebox{0.75}{$\ketbra{\nu_{1}}{\nu_{2}}+$}\frac{\rho_{21}\sqrt{\lambda_{1}\lambda_{2}}}{|\rho_{21}|}\scalebox{0.75}{$\ketbra{\nu_{2}}{\nu_{1}}+\lambda_{1}\proj{\nu_{2}}$}
\Big],\\
\bar{M}_{1}^{\rm opt}&=&\frac{\eta_{1}}{\lambda_{1}+\lambda_{2}+2q-1-C_{1}+C_{2}}
\Big[\scalebox{0.75}{$(\lambda_{2}+q-1+C_{2})\proj{\nu_{1}}
+$}\frac{\rho_{12}\sqrt{\lambda_{1}\lambda_{2}}}{|\rho_{12}|}\scalebox{0.75}{$\ketbra{\nu_{1}}{\nu_{2}}$}\\
&&\hfill+\frac{\rho_{21}\sqrt{\lambda_{1}\lambda_{2}}}{|\rho_{21}|}\scalebox{0.75}{$\ketbra{\nu_{2}}{\nu_{1}}+(\lambda_{1}+q-C_{1})\proj{\nu_{2}}$}
\Big],\\
\bar{M}_{2}^{\rm opt}&=&\frac{\eta_{2}}{\lambda_{1}+\lambda_{2}+2q-1+C_{1}-C_{2}}\Big[
\scalebox{0.75}{$(\lambda_{2}+q-C_{2})\proj{\nu_{1}}
+$}\frac{\rho_{12}\sqrt{\lambda_{1}\lambda_{2}}}{|\rho_{12}|}\scalebox{0.75}{$\ketbra{\nu_{1}}{\nu_{2}}$}\\
&&\hfill+\frac{\rho_{21}\sqrt{\lambda_{1}\lambda_{2}}}{|\rho_{21}|}\scalebox{0.75}{$\ketbra{\nu_{2}}{\nu_{1}}+(\lambda_{1}+q-1+C_{1})\proj{\nu_{2}}$}\Big].
\end{array}
\end{equation}
Here $\eta_{0}$ becomes $P_{\rm I}(q)$ of \eqref{eq:pcpi2}, and
$\eta_{1}$ and $\eta_{2}$ are given by
\begin{equation}\label{eq:eta12}
\begin{array}{rll}
\eta_{1}&=&
\Big[ \frac{(2C_{1}-1)(2C_{2}-1)+2(C_{2}-C_{1})(2q-1)+(2q-1)^{2}}{2(2q-1)^{2}(C_{1}+C_{2}-1)} \Big]\!\cdot\!
\Big[\scalebox{0.75}{$\rho_{11}(q-1+C_{1})+\rho_{22}(C_{2}-q)$}\\
&& -\frac{|\rho_{12}|(C_{2}-q)(q-1+C_{1})[(2C_{1}-1)(q-1+C_{2})+(2C_{2}-1)(C_{1}-q)]}{\sqrt{(2C_{1}-1)(2C_{2}-1)(C_{1}-q)(C_{2}-q)(q-1+C_{1})(q-1+C_{2})}}\Big],\\
\eta_{2}&=&
\Big[ \frac{(2C_{1}-1)(2C_{2}-1)-2(C_{2}-C_{1})(2q-1)+(2q-1)^{2}}{2(2q-1)^{2}(C_{1}+C_{2}-1)} \Big]\!\cdot\! \Big[\scalebox{0.75}{$\rho_{11}(C_{1}-q)+\rho_{22}(q-1+C_{2})$}\\
&&-\frac{|\rho_{12}|(C_{1}-q)(q-1+C_{2})[(2C_{2}-1)(q-1+C_{1})+(2C_{1}-1)(C_{2}-q)]}{\sqrt{(2C_{1}-1)(2C_{2}-1)(C_{1}-q)(C_{2}-q)(q-1+C_{1})(q-1+C_{2})}}\Big].
\end{array}
\end{equation}
Two cases are distinguished by the signs of $\{\lambda_{i}\}_{i=1}^{2}$ and
$\{\eta_{i}\}_{i=1}^{3}$. If $\lambda_{1},\lambda_{2}\geq0$ and
$\eta_{1},\eta_{2},\eta_{3}>0$, we have $M_{i}^{\rm opt}\neq
0(\forall i)$. Otherwise, we obtain $M_{1}^{\rm opt}=0$ or
$M_{2}^{\rm opt}=0$.
\end{theorem}
The proof is given in Appendix \ref{app:the3}. In Theorem \ref{the:q0q1m}, optimal POVM corresponding to $q_{0}=q$ is always unique and $P_{\rm I}(q)$ becomes a set with only one element. In this case, we consider $P_{\rm I}(q)$ as a value corresponding to the element of the set, like \eqref{eq:pcpi1} and \eqref{eq:pcpi2}.\\
\indent The following lemma shows the result related with inconclusive degrees
satisfying $M_{1}^{\rm opt}=0$.
\begin{lemma}\label{lem:m10}
When $\rho_{12}\neq0$, if $C_{1}\leq\frac{1}{2}<C_{2}$ and
$q_{0}^{(0)}<q_{0}<q_{0}^{(1)}$(or $\frac{1}{2}<C_{1}\leq C_{2}$ and
$C_{1}<q_{0}<q_{0}^{(1)}$) can be satisfied, the optimal POVM element
$M_{1}^{\rm opt}$ of modified FRIR problem becomes 0.
\end{lemma}
The proof is given in Appendix \ref{app:lem3}.\\
\indent To obtain $R_{\rm cor}^{\rm opt}(Q)$, we need to express the
inconclusive degree as a function of the failure probability. This
task is not easy since the relation between inconclusive degree and
failure probability is very complex; see $P_{\rm I}(q)$ of
\eqref{eq:pcpi1} and \eqref{eq:pcpi2}. However, it should be noted
that the relation between $q$ and $P_{\rm I}(q)$ is one-to-one,
which implies that we can obtain $R_{\rm cor}^{\rm opt}(Q)$
numerically using Eq. \eqref{eq:rbarp} and theorem \ref{the:q0q1m}.\\
\indent For example, let us consider the following case of
$\{q_{i},\rho_{i}\}_{i=1}^{2}$.
\begin{eqnarray}
q_{1}=0.4,\,
\rho_{1}=\left(
\begin{array}{cc}
0.15&-0.30+0.10i\\
-0.30-0.10i&0.85
\end{array}
\right),\nonumber\\
q_{2}=0.6,\,
\rho_{2}=\left(
\begin{array}{cc}
0.80&-0.30+0.05i\\
-0.30-0.05i&0.20
\end{array}
\right).
\end{eqnarray}
The Bloch vectors of the two qubit states are 
\begin{equation*}
{\bm v}_{1}=(-0.6,-0.2,-0.7)\quad\mbox{and}\quad{\bm v}_{2}=(-0.6,-0.1,0.6). 
\end{equation*}
Then,
$|\rho_{12}|$, $C_{1}$, and $C_{2}$ become $0.3075$, $0.8361$, and
$0.9657$, respectively. From corollary \ref{cor:q01} and theorem
\ref{the:q00}, $q_{0}^{(0)}$ and $q_{0}^{(1)}$ are $\chi=0.6940$ and
$C_{2}$, respectively. In the region of $\chi<q\leq C_{1}$,
$\lambda_{1}$ and $\lambda_{2}$ are non-negative, and $\eta_{0}$ and
$\eta_{2}$ are positive, but $\eta_{1}$ is not. In $\chi <q<0.7902$,
$\eta_{1}$ is positive. However, in $0.7902\leq q \leq C_{1}$,
$\eta_{1}$ is negative or equal to zero. Therefore, in $\chi
<\!q_{0}\!<0.7902$, we find $M_{i}^{\rm opt}\neq 0(\forall i)$. However, in
$0.7902\leq\!q_{0}\!\leq C_{1}$, because of $q_{1}+\|q_{0}{\bm
v}_{0}-q_{1}{\bm v}_{1}\|_{2}<q_{2}+\|q_{0}{\bm v}_{0}-q_{2}{\bm
v}_{2}\|_{2} $, we have $M_{1}^{\rm opt}\!=\!0$. In addition, by lemma
\ref{lem:m10}, in $C_{1}<q_{0}<C_{2}$, $M_{1}^{\rm opt}$ becomes 0.
Therefore if fixed failure probability $P_{\rm I}$ is $0<P_{\rm
I}<0.5805$, we find $M_{i}^{\rm opt}\neq 0$. However, if $0.5805\leq
P_{\rm I}<Q_{1}=0.6635$, we have $M_{0}^{\rm opt}\neq 0$,
$M_{1}^{\rm opt}=0$, and $M_{2}^{\rm opt}\neq 0$. Figure
\ref{fig:ex} shows, in this example, the behavior of $P_{\rm
I}(q_{0})$ and $\bar{P}_{\rm cor}^{\rm opt}(q)$($R_{\rm cor}^{\rm
opt}(Q)$) in
$q_{0}^{(0)}<q_{0}<q_{0}^{(1)}$($0<Q<1$).\\
\indent To confirm the effectiveness of our results, we consider the
case of $C_{1}=C_{2}(=C)$ and $\rho_{12}\neq 0$. By corollary
\ref{cor:q01} and theorem \ref{the:q00}, in this case,
$q_{0}^{(0)}=\chi$ and $q_{0}^{(1)}=C$. If $C_{1}=C_{2}$ is applied to
$\lambda_{1}$,\,$\lambda_{2}$,\,$\eta_{0}$,\,$\eta_{1}$,\,and
$\eta_{2}$, we have the following expresssion.
\begin{eqnarray}\label{eq:epp}
\lambda_{1}=\lambda_{2}=\frac{(C-q)(q-1+C)}{2q-1},
\end{eqnarray}
and
\begin{equation}\label{eq:epp}
\begin{array}{rll}
\eta_{0}&=&\frac{2}{(2q-1)^{2}}\Big[\scalebox{0.75}{$
|\rho_{12}|(C-q)^{2}+|\rho_{12}|(q-1+C)^{2}-(C-q)(q-1+C)$}\Big],\\
\eta_{1}&=&\frac{(2C-1)^{2}+(2q-1)^{2}}{2(2q-1)^{2}(2C-1)}\Big[\scalebox{0.75}{$
\rho_{11}(q-1+C)+\rho_{22}(C-q)-|\rho_{12}|(2C-1)$}\Big],\\
\eta_{2}&=&\frac{(2C-1)^{2}+(2q-1)^{2}}{2(2q-1)^{2}(2C-1)}\Big[\scalebox{0.75}{$
\rho_{11}(C-q)+\rho_{22}(q-1+C)-|\rho_{12}|(2C-1)$}\Big].
\end{array}
\end{equation}
When $\rho_{11},\rho_{22}\geq|\rho_{12}|$, in $\chi<q<C$, these
are all positive, and we have $M_{i}^{\rm opt}\neq0(\forall i)$.
Then $q_{0}(=q)$ can be expressed in terms of the failure
probability $Q$:
\begin{eqnarray}\label{eq:epp}
q=
\frac{1}{2} +\frac{2C-1}{2}\sqrt{\frac{1-2|\rho_{12}|}{1+2|\rho_{12}|-2Q}}.
\end{eqnarray}
Applying this $q$ to $\bar{P}_{\rm cor}^{\rm opt}(q)-qQ$, we find
$P_{\rm cor}^{\rm opt}(Q)$ in $0<Q<2|\rho_{12}|$, which agrees with
the previous result:
\begin{eqnarray}\label{eq:ccop}
P_{\rm cor}^{\rm opt}(Q)
=\frac{1-Q}{2}+\frac{2C-1}{2}\sqrt{(1-2|\rho_{12}|)(1+2|\rho_{12}|-2Q)}.
\end{eqnarray}
When $\rho_{11}<|\rho_{12}|\leq\rho_{22}$, then
$\lambda_{1},\lambda_{2},\eta_{0}$,\,and $\eta_{2}$ are always positive in
the region of $\chi<q<C$; however, $\eta_{1}$ can be found only in the following case\\
\begin{eqnarray}
\chi <q
<\frac{1}{2}+\frac{(1-2|\rho_{12}|)(2C-1)}{2(\rho_{22}-\rho_{11})}.
\end{eqnarray}
In this region, like \eqref{eq:epp}, $q$ can be expressed by
$Q$. Therefore, in the following region of the failure probability,
$P_{\rm cor}^{\rm opt}(Q)$ is the same as \eqref{eq:ccop}, and we
have $M_{i}^{\rm opt}\neq0(\forall i)$.
\begin{eqnarray}
0<Q<\frac{2(\rho_{11}\rho_{22}-|\rho_{12}|^{2})}{1-2|\rho_{12}|}.
\end{eqnarray}
In the other region, we get $M_{0}^{\rm opt}\neq0,M_{1}^{\rm
opt}=0$, and $M_{2}^{\rm opt}\neq 0$, which
coincides with the previous result\cite{ref:herz3}.
\section{Conclusion}\label{sec:con}
In this paper, we provided a solution to the FRIR of two mixed qubit
states. The solution was obtaind by considering the modified FRIR
problem(MD of three qubit states). In fact, since the added specific
quantum state $\rho_{0}$ with the prior probability $q_{0}$(called
an conclusive degree) was obtained from the given two qubit states,
the structure of the modified problem is more complex than that of
the MD of three qubit states with no constraint\cite{ref:ha1}.
First, we introduced special inconclusive degrees $q_{0}^{(0)}$ and
$q_{0}^{(1)}$, which are the beginning and the end of proper
inconclusive degrees. Using this, we divided the problem into the
two cases of $q_{0}=q_{0}^{(0)}$(or $q_{0}=q_{0}^{(1)}$) and
$q_{0}^{(0)}<q_{0}<q_{0}^{(1)}$. By maximum confidences of two qubit
states and non-diagonal element of $\rho_{0}$, we solved each case.
We obtained $q_{0}^{(0)}$ and $q_{0}^{(1)}$ in the analytic form,
and completely understood modified FRIR problem when
$q_{0}^{(0)}\leq q_{0}\leq q_{0}^{(1)}$. Finally, we verified that
our results also provide the same solutions as known examples in the
literature.
\begin{acknowledgements}
This work is supported by the Basic Science Research Program through
the National Research Foundation of Korea funded by the Ministry of
Education, Science and Technology(NRF2015R1D1A1A01060795) and
Institute for Information \& communications Technology
Promotion(IITP) grant funded by the Korea government(MSIP)(No. R0190-15-2028, PSQKD).
\end{acknowledgements}
\appendix
\section{Proofs of Lemmas in Section \ref{sec:FRIR}}\label{app:lem2}
\noindent{\bf Proof of Lemma \ref{lem:rel}} Suppose that when $\{q_{i},\rho_{i}\}_{i=1}^{N}$
is given, POVM $\{M_{i}\}_{i=0}^{N}$ can cause $P_{\rm I}=Q$ and $\bar{P}_{\rm cor}=\bar{P}_{\rm
cor}^{\rm opt}(q)$. It follows that the
POVM can make $P_{\rm cor}(Q)=\bar{P}_{\rm cor}^{\rm opt}(q)-qQ$. If
there exists a POVM that can build $P_{\rm I}=Q$ and $P_{\rm cor}(Q)\!=\!P\!>\!\bar{P}_{\rm
cor}^{\rm opt}(q)-qQ$, it can also
construct $\bar{P}_{\rm cor}(q)\!=\!P+qQ$.
However since $P+qQ$ is larger than $\bar{P}_{\rm cor}^{\rm opt}(q)$, this is contradictory.
Therefore $\{M_{i}\}_{i=0}^{N}$ should produce $P_{\rm cor}^{\rm opt}(Q)$,
which means $P_{\rm cor}^{\rm opt}(Q)=\bar{P}_{\rm cor}^{\rm opt}(q)-qQ$.\hfill$\square$\\

\noindent{\bf Proof of Lemma \ref{lem:con}} Assume that when $\{q_{i},\rho_{i}\}_{i=1}^{N}$
is given, the POVM $\{M_{i}\}_{i=0}^{N}$($\{M_{i}'\}_{i=0}^{N}$) can
produce $P_{\rm I}=Q$ and $\bar{P}_{\rm cor}=\bar{P}_{\rm cor}^{\rm opt}(q)$($P_{\rm I}=Q'$ and $\bar{P}_{\rm cor}=\bar{P}_{\rm cor}^{\rm opt}(q')$). If
$q=q'$ and $Q<Q'$, the POVM $\{M_{i}''\}_{i=0}^{N}$ composed of
$M_{i}''=pM_{i}+(1-p)M_{i}'$($0\leq p \leq 1$) will build
$\bar{P}_{\rm cor}=\bar{P}_{\rm cor}^{\rm opt}(q)$
and $P_{\rm I}=pQ+(1-p)Q'$. Therefore $P_{\rm I}(q)$ becomes a convex set.\\
\indent Now suppose that $q<q'$. $\{M_{i}\}_{i=0}^{N}$ constructs $\bar{P}_{\rm
cor}=(q'-q)Q+\bar{P}_{\rm cor}^{\rm opt}(q)$ when $q_{0}=q'$, and the value
should be equal to or less than $\bar{P}_{\rm cor}^{\rm
opt}(q')=(q'-q)Q'+\bar{P}$, where $\bar{P}$ is $\bar{P}_{\rm cor}$ corresponding to $\{M_{i}'\}_{i=0}^{N}$ when $q_{0}=q$.
This means that
$(Q'-Q)\geq(\bar{P}_{\rm cor}^{\rm opt}(q)-\bar{P})/(q'-q)$.
Therefore we have $Q\leq Q'$. This means that $P_{\rm I}(q)\leq P_{\rm I}(q')$.\hfill$\square$\\

\noindent{\bf Proof of Lemma \ref{lem:eq}.} When $q_{0}<1/N$, we get
$\bar{\tau}_{0}^{\rm opt}=(1/N-q_{0})I_{d}+(1/N)\sum_{i=1}^{N}\bar{\tau}_{i}^{\rm
opt}$ by (ii)
of \eqref{eq:optc1}. If we multiply $\bar{M}_{0}^{\rm
opt}$ to both sides of the equation and take the trace of the
result, we obtain $(1/N-q_{0})\tr[\bar{M}_{0}^{\rm opt}]\leq 0$ by
(iii) and the positivity of $\bar{\tau}_{i}^{\rm
opt}$($\forall i$). From the assumption on $q_{0}$, $\bar{M}_{0}^{\rm
opt}$ should be zero and we find $P_{\rm I}(q)=0(\forall q<1/N)$.
Therefore using lemma \ref{lem:con}, we have $q_{0}^{(0)}\geq 1/N$.\\
\indent  When $C=\max_{i}C_{i}$, $\bar{M}_{i}=\delta_{i0}\rho_{0}$ and $\bar{\tau}_{i}=CI_{d}-\bar{\rho}_{i}$ satisfy the optimality condition \eqref{eq:optc1} of $q_{0}=C$,
and $1\in P_{\rm I}(C)$. Therefore we get $q_{0}^{(1)}\leq C$ by lemma \ref{lem:con}.\hfill$\square$
\section{Proofs of Lemmas in Section \ref{sec:two}}\label{app:lem3}
\noindent{\bf Proof of Lemma \ref{lem:q01}.}
When $q_{0}=C_{2}$, since $\{\bar{M}_{i}=\delta_{i0}\rho_{0}\}_{i=0}^{2}$ and
$\{\bar{\tau}_{i}^{\star}=C_{2}I_{2}-\bar{\rho}_{i}\}_{i=0}^{2}$ satisfy
the KKT optimality condition \eqref{eq:optc1},
$\bar{P}_{\rm cor}^{\rm opt}(C_{2})=C_{2}$.
This means that $\bar{\tau}_{i}^{\rm opt}=\bar{\tau}_{i}^{\star}(\forall i)$.
Since the rank of $\bar{\tau}_{2}^{\rm opt}$ should be
one by $C_{1}+C_{2}>1$, we
have $\bar{M}_{2}^{\rm opt}=\beta\proj{\nu_{2}}$ from (iii). The form of
$\bar{M}_{0}^{\rm opt}$ and $\bar{M}_{1}^{\rm opt}$ can be
classified into the cases of $C_{1}=C_{2}$ and $C_{1}<C_{2}$.\\
\indent If $C_{1}=C_{2}$, the rank of $\bar{\tau}_{1}^{\rm opt}$
becomes 1, and we get $\bar{M}_{1}^{\rm opt}=\alpha\proj{\nu_{1}}$
from (iii). Furthermore (i) indicates that $\bar{M}_{0}^{\rm
opt}=\rho_{0}-\alpha\proj{\nu_{1}}-\beta\proj{\nu_{2}}$,
$0\leq\alpha\leq\rho_{11}$, $0\leq\beta\leq\rho_{22}$, and
$(\rho_{11}-\alpha)(\rho_{22}-\beta)\geq|\rho_{12}|^{2}$. Since
$\tr[\bar{M}_{0}^{\rm opt}]=1-\alpha-\beta$, the maximum can be
found at $\alpha=\beta=0$. $\alpha$ and $\beta$ corresponding to its
minimum can be different. When $\rho_{11}<|\rho_{12}|\leq\rho_{22}$,
we have $\alpha=0$ and $\beta=1-Q_{1}$. When
$\rho_{22}<|\rho_{12}|\leq\rho_{11}$, we obtain $\alpha=1-Q_{2}$ and
$\beta=0$. When $|\rho_{12}|\leq\rho_{11},\rho_{22}$, we have
$\alpha=\rho_{11}-|\rho_{12}|$ and $\beta=\rho_{22}-|\rho_{12}|$.
Therefore $P_{\rm I}(C_{2})$ becomes \eqref{eq:q01pi}.\\
\indent If $C_{1}<C_{2}$, the rank of  $\bar{\tau}_{1}^{\rm opt}$
becomes 2, and (iii) implies that $\bar{M}_{1}^{\rm opt}=0$. Then
(i) means $\bar{M}_{0}^{\rm opt}=\rho_{0}-\beta\proj{\nu_{2}}$ and
$0\leq\beta\leq 1-Q_{1}$. Since $\tr[\bar{M}_{0}^{\rm
opt}]=1-\beta$, the minimum(maximum) can be found at
$\beta\!=\! 1-Q_{1}$($\beta=0$). Therefore we get $P_{\rm
I}(C_{2})\!=\![Q_{1},1]$.
\hfill$\square$\\

\noindent{\bf Proof of Lemma \ref{lem:q001}.} When $C_{1}\leq 1/2<C_{2}$, since
$\{\bar{M}_{i}=\delta_{i2}\rho_{0}\}_{i=0}^{2}$ and the following
$\{\bar{\tau}_{i}^{\star}\}_{i=0}^{2}$ satisfies the optimality
condition \eqref{eq:optc1} of $q_{0}=1-C_{1}$, $\bar{P}_{\rm
cor}^{\rm opt}(1-C_{1})=q_{2}$.
\begin{eqnarray}
\bar{\tau}_{0}^{\star}=(C_{1}+C_{2}-1)\proj{\nu_{2}},\
\bar{\tau}_{1}^{\star}=(1-2C_{1})\proj{\nu_{1}}+(2C_{2}-1)\proj{\nu_{2}},\
\bar{\tau}_{2}^{\star}=0.
\end{eqnarray}
This implies that $\bar{\tau}_{i}^{\rm opt}=\bar{\tau}_{i}^{\star}(\forall i)$ when $q_{0}=1-C_{1}$.
Since the rank of $\bar{\tau}_{0}^{\rm opt}$ is 1 given that $C_{1}+C_{2}>1$, (iii) implies $\bar{M}_{0}^{\rm opt}=\alpha\proj{\nu_{1}}$. However
$\bar{M}_{1}^{\rm opt}$ and $\bar{M}_{2}^{\rm opt}$  are classified
into cases where $C_{1}<1/2$ and $C_{1}=1/2$.\\
\indent When $C_{1}<1/2$,
since the rank of $\bar{\tau}_{1}^{\rm opt}$ becomes 2, (iii) gives $\bar{M}_{1}^{\rm opt}=0$ and (i) means $\bar{M}_{2}^{\rm
opt}=\rho_{0}-\alpha\proj{\nu_{1}}$ and $0\leq\alpha\leq 1-Q_{2}$. Then,
since $\tr[\bar{M}_{0}^{\rm opt}]=\alpha$ shows a minimum at
$\alpha=0$ and a maximum at $\alpha=1-Q_{2}$, we have $P_{\rm
I}(1-C_{1})=[0,1-Q_{2}]$. However, when $C_{1}=1/2$, the rank of
$\bar{\tau}_{1}^{\rm opt}$ is 1 and (iii) implies $\bar{M}_{1}^{\rm
opt}=\beta\proj{\nu_{1}}$. Therefore (i) means $\bar{M}_{2}^{\rm opt}=\rho_{0}-(\alpha+\beta)\proj{\nu_{1}}$
and we have $\alpha,\beta\geq 0$, and $\alpha+\beta\leq 1-Q_{2}$.
$\tr[\bar{M}_{0}^{\rm opt}]=\alpha$ has a minimum(maximum) at $\alpha=0$($\alpha=1-Q_{2}$ and $\beta=0$).
Therefore, we obtain $P_{\rm I}(1-C_{1})=[0,1-Q_{2}]$.\hfill$\square$\\

\noindent{\bf Proof of Lemma \ref{lem:q002}.}
When $1/2<C_{1}\leq C_{2}$,$\rho_{12}=0$, since
the following $\{\bar{M}_{i}\}_{i=0}^{2}$,\,$\{\bar{\tau}_{i}^{\star}\}_{i=0}^{2}$
satisfies the optimality condition \eqref{eq:optc1} of $q_{0}=C_{1}$, $\bar{P}_{\rm cor}^{\rm
opt}(C_{1})=\rho_{11}C_{1}+\rho_{22}C_{2}$.
\begin{alignat}{3}
\bar{M}_{0}&=0,\,
&\bar{M}_{1}&=\rho_{11}\proj{\nu_{1}},\,
&\bar{M}_{2}&=\rho_{22}\proj{\nu_{2}},\nonumber\\
\bar{\tau}_{0}^{\star}&=(C_{2}-C_{1})\proj{\nu_{2}},\,
&\bar{\tau}_{1}^{\star}&=(2C_{2}-1)\proj{\nu_{2}},\,
&\bar{\tau}_{2}^{\star}&=(2C_{1}-1)\proj{\nu_{1}}.
\end{alignat}
This means that $\bar{\tau}_{i}^{\rm
opt}=\bar{\tau}_{i}^{\star}(\forall i)$ when $q_{0}=C_{1}$. Since,
if $C_{1}<C_{2}$, the rank of $\bar{\tau}_{i}^{\rm opt}$ is one,
(iii) tells that $\bar{M}_{0}^{\rm opt}$ and $\bar{M}_{1}^{\rm opt}$
are proportional to $\proj{\nu_{1}}$, and $\bar{M}_{2}^{\rm opt}$ is
proportional to $\proj{\nu_{2}}$. By (i), $\bar{M}_{i}^{\rm opt}$
can be expressed as \eqref{eq:m31}. However, since $C_{1}=C_{2}$
implies $\bar{\tau}_{0}^{\rm opt}=0$, $\bar{M}_{i}^{\rm opt}$
becomes \eqref{eq:m32}.
Therefore $P_{\rm I}(C_{1})$ can be written as $[0, \rho_{11}+\rho_{22}\delta_{C_{1},C_{2}}]$. \hfill$\square$\\

\noindent{\bf Proof of Lemma \ref{lem:q004}.} When $1/2<C_{1}\leq C_{2}$ and $\rho_{12}\neq 0$, lemma
\ref{lem:con} and corollary \ref{cor:q01} reveal that
$q_{0}^{(0)}<C_{2}$. By (ii) of optimality condition
\eqref{eq:optc1} and the nonnegativity of $\bar{\tau}_{i}(\forall
i)$, $\bar{\tau}_{0}^{\rm opt}=0$ includes $q_{0}\geq C_{2}$, and
$\bar{\tau}_{1(2)}^{\rm opt}=0$ contains $C_{2(1)}\leq 1/2$. This
implies that $\bar{\tau}_{i}^{\rm opt}\neq 0(\forall i)$ if
$q_{0}<q_{0}^{(1)}$. Then $\bar{M}_{0}^{\rm opt}=0$ means that
$\bar{M}_{1}^{\rm opt},\bar{M}_{2}^{\rm opt}\neq 0$ because
$\bar{M}_{0}^{\rm opt}=\bar{M}_{1}^{\rm opt}=0$ implies
$\bar{\tau}_{2}^{\rm opt}=0$ and $\bar{M}_{0}^{\rm
opt}=\bar{M}_{2}^{\rm opt}=0$ contains $\bar{\tau}_{1}^{\rm opt}=0$.
When $\bar{M}_{0}^{\rm opt}=0$, in order to obtain the explicit form
of $\{\bar{M}_{i}^{\rm opt},\bar{\tau}_{i}^{\rm opt}\}_{i=1}^{2}$,
we use the optimality condition \eqref{eq:optc2}. Since
$\bar{M}_{0}=0$ includes $p_{0}=0$, it has no effect on $r_{0}$ and
${\bm w}_{0}$. $\bar{M}_{i},\bar{\tau}_{i}\neq 0(i=1,2)$ implies
$p_{i},r_{i}\neq 0 (i=1,2)$, and by (iii) we have $\|{\bm
w}_{i}\|_{2}=1$, ${\bm u}_{i}=-{\bm w}_{i}(i=1,2)$. Since
$r_{1}+r_{2}=l$ and $r_{2}-r_{1}=q_{1}-q_{2}$ should be satisfied by
(ii), $\{p_{i}^{\rm opt},{\bm u}_{i}^{\rm opt}\}_{i=1}^{2}$ and
$\{r_{i}^{\rm opt},{\bm w}_{i}^{\rm opt}\}_{i=1}^{2}$ can be
obtained as follows:
\begin{eqnarray}\label{eq:pruw}
p_{1}^{\rm opt}=
p_{2}^{\rm opt}=\frac{1}{2},\,
r_{1}^{\rm opt}=\frac{1+l}{2}-q_{1},\,
r_{2}^{\rm opt}=\frac{1+l}{2}-q_{2},\nonumber\\
{\bm u}_{1}^{\rm opt}={\bm w}_{2}^{\rm opt}=\frac{q_{1}{\bm v}_{1}-q_{2}{\bm v}_{2}}{\|q_{1}{\bm v}_{1}-q_{2}{\bm v}_{2}\|_{2}},\,
{\bm u}_{2}^{\rm opt}={\bm w}_{1}^{\rm opt}=\frac{q_{2}{\bm v}_{2}-q_{1}{\bm v}_{1}}{\|q_{1}{\bm v}_{1}-q_{2}{\bm v}_{2}\|_{2}}.
\end{eqnarray}
From these, we find $\bar{P}_{\rm cor}^{\rm opt}(q)=(1+l)/2(\forall
q\leq q_{0}^{(0)})$, and can decide the explicit form of
$\{{M}_{i}^{\rm opt}\}_{i=0}^{2}$ and $\{\tau_{i}^{\rm
opt}\}_{i=1}^{2}$. However $r_{0}^{\rm opt}$ and ${\bm w}_{0}^{\rm
opt}$ are not decided yet. These are affected only by (ii). The
triangle made of $\{q_{i}{\bm v}_{i}\}_{i=0}^{2}$ lies in the plane
with the origin, and the triangle consisting of $\{-r_{i}^{\rm
opt}{\bm w}_{i}^{\rm opt}\}_{i=0}^{2}$ should be located in the same
plane. Since the two triangles are congruent, then as $\|r_{0}^{\rm
opt}{\bm w}_{0}^{\rm opt}\|_{2}$ grows larger $\|q_{0}{\bm
v}_{0}\|_{2}$ becomes larger. Since $q_{0}+r_{0}^{\rm opt}$ is fixed
as $(1+l)/2$, when $\|{\bm w}_{0}^{\rm opt}\|_{2}$ reaches the
maximum(that is, when $\|{\bm w}_{0}^{\rm opt}\|_{2}=1$), $q_{0}$
reaches the maximum. Therefore the determinant of
$\bar{\tau}_{0}^{\rm opt}$ is 0 when $q_{0}=q_{0}^{(0)}$. From (ii),
we have
$(\chi_{1}-q_{0}^{(0)})(\chi_{2}-q_{0}^{(0)})=|\gamma_{12}|^{2}$.
Though there are two roots of this equation, the nonnegativity of
$\bar{\tau}_{0}^{\rm opt}$ implies that $q_{0}^{(0)}\leq
\min\{\chi_{1},\chi_{2}\}$, and the analytic form of $q_{0}^{(0)}$
can be obtained as $\chi$ of \eqref{eq:chi}. The optimal POVM of $q_{0}=\chi$ is unique since $r_{i}^{\rm opt}\neq 0(\forall i)$ and $\{q_{i}{\bm v}_{i}\}_{i=0}^{2}$ forms a triangle; see the Appendix \ref{app:ano}. This means that ${P}_{\rm I}(\chi)=0$.\hfill$\square$\\

\noindent{\bf Proof of Lemma \ref{lem:uni}.} When $C_{1}<C_{2}$ and $\rho_{12}=0$, if
$q_{0}^{(0)}<q<q_{0}^{(1)}$, POVM, defined as \eqref{eq:m41}, and
$\{\tau_{i}^{\star}\}_{i=0}^{2}$ satisfies KKT optimality condition
\eqref{eq:optc1} to $q_{0}=q$:
\begin{eqnarray}
\bar{\tau}_{0}^{\star}&=&(C_{2}-q)\proj{\nu_{2}},\nonumber\\
\bar{\tau}_{1}^{\star}&=&(q-C_{1})\proj{\nu_{1}}+(2C_{2}-1)\proj{\nu_{2}},\nonumber\\
\bar{\tau}_{2}^{\star}&=&(q-1+C_{1})\proj{\nu_{1}}.
\end{eqnarray}
This means that $\bar{\tau}_{i}^{\rm
opt}=\bar{\tau}_{i}^{\star}(\forall i)$. Since the rank of
$\bar{\tau}_{1}^{\rm opt}$ is two, (iii) implies $\bar{M}_{1}^{\rm
opt}=0$. However, since the rank of $\bar{\tau}_{0}^{\rm opt}$ and
$\bar{\tau}_{2}^{\rm opt}$ are one, $\bar{M}_{0}^{\rm opt}$ and
$\bar{M}_{2}^{\rm opt}$ are proportional to $\proj{\nu_{1}}$ and
$\proj{\nu_{2}}$, respectively. Therefore (i) means that
$\bar{M}_{i}^{\rm opt}$ is unique as \eqref{eq:m41}.
\hfill$\square$\\

\noindent{\bf Proof of Lemma \ref{lem:m10}.} First of all, let us consider the case of
$C_{1}\leq\frac{1}{2}<C_{2}$ and $\rho_{12}\neq0$.
 In the region of $q_{0}^{(0)}<q_{0}<q_{0}^{(1)}$, since $\lambda_{1}$ of \eqref{eq:lam12} is less than 0, we find $M_{1}^{\rm opt}=0$ or $M_{2}^{\rm opt}=0$ by theorem \ref{the:q0q1m}.
If $M_{2}^{\rm opt}=0$, since optimality condition \eqref{eq:optc1}
means $\bar{\tau}_{1}^{\rm opt}=\bar{\tau}_{0}^{\rm
opt}+q_{0}I_{2}-\bar{\rho}_{1}$ and
  $\det(\bar{\tau}_{0}^{\rm opt})=\det(\bar{\tau}_{1}^{\rm opt})=0$,
$t_{i}=\bra{\nu_{i}}\bar{\tau}_{0}^{\rm opt}\ket{\nu_{i}}$ satisfies
$t_{1}t_{2}=(t_{1}+q_{0}-C_{1})(t_{2}+q_{0}-1+C_{2})$. However, this
result is contradictory because $(t_{1}+q_{0}-C_{1})(t_{2}+q_{0}-1+C_{2})$ is greater than $t_{1}t_{2}$ in the region of $(C_{1},1-C_{2}\leq 1-C_{1}=)q_{0}^{(0)}<q_{0}<q_{0}^{(1)}$. Therefore we get $M_{1}^{\rm opt}=0$.\\
\indent Next, let us consider the case of $\frac{1}{2}<C_{1}\leq C_{2}$ and
$\rho_{12}\neq0$. Here $(q_{0}^{(0)},q_{0}^{(1)})$ is divided into
two cases: $(\chi,C_{1}]$ and $(C_{1},C_{2})$. In latter case,
because of $\lambda_{1}< 0$, we can obtain $M_{1}^{\rm opt}=0$.\hfill$\square$
\section{Proof of Theorem \ref{the:q0q1m}}\label{app:the3}
\noindent {\bf Proof.} When $\rho_{xy}\neq 0$ and
$q_{0}=q\in(q_{0}^{(0)},q_{0}^{(1)})$, the line intersecting ${\bm v}_{1}$
and ${\bm v}_{2}$ does not contain the origin, and $\{q_{i}{\bm
v}_{i}\}_{i=0}^{2}$ forms a triangle. $r_{0}^{\rm opt}=0$ implies
that $\{M_{i}=\delta_{i0}I_{2}\}_{i=0}^{2}$ provide an optimal POVM,
which includes $q_{0}^{(1)}\leq q$. Since $r_{k}^{\rm
opt}=0(k\in\{1,2\})$ indicates that
$\{M_{i}=\delta_{ik}I_{2}\}_{i=0}^{2}$ yields the optimal POVM, this
means $q\leq q_{0}^{(0)}$. Therefore the element of
$\{r_{i}^{\rm opt}\}_{i=0}^{2}$ are all nonzero. In this case, the
optimal POVM is unique; see the Appendix \ref{app:ano}. In addition, $M_{0}^{\rm opt}$ is nonzero,
and at least one of $M_{1}^{\rm opt}$ and $M_{2}^{\rm opt}$ is nonzero.\\
\indent In the case of $M_{x}^{\rm opt}\neq0$,\,$M_{y}^{\rm opt}=0(\{x,y\}=\{1,2\})$, the index $x$ turns out to be the
index $i$ in $\max_{i\in{1,2}}[q_{i}+\|q{\bm v}_{0}-q_{i}{\bm v}_{i}\|_{2}]$ because $\bar{P}_{\rm cor}^{\rm opt}(q)=\max_{i\in{1,2}}[q+q_{i}+\|q\rho_{0}-q_{i}\rho_{i}\|_{1}]/2$. The optimal POVM, by the optimality condition \eqref{eq:optc2}, can
be expressed as \eqref{eq:optm1}.\\
\indent In the case of $M_{i}^{\rm opt}\neq 0(\forall i)$, by the
optimality condition \eqref{eq:optc1}, $\{\bar{M}_{i}^{\rm
opt},\bar{\tau}_{i}^{\rm opt}\}_{i=0}^{2}$ can be found explicitly.
From condition (ii), $\{\bar{\tau}_{i}^{\rm opt}\}_{i=0}^{2}$ are
given as follows.
\begin{eqnarray}\label{eq:tau}
 \bar{\tau}_{0}^{\rm opt}&=&\tau_{11}\ketbra{\nu_{1}}{\nu_{1}}+\tau_{12}\ketbra{\nu_{1}}{\nu_{2}}+\tau_{21} \ketbra{\nu_{2}}{\nu_{1}}+\tau_{22}\ketbra{\nu_{2}}{\nu_{2}},\nonumber\\
 \bar{\tau}_{1}^{\rm opt}&=&(\tau_{11}+q-C_{1})\ketbra{\nu_{1}}{\nu_{1}}+\tau_{12}\ketbra{\nu_{1}}{\nu_{2}}+\tau_{21} \ketbra{\nu_{2}}{\nu_{1}}+(\tau_{22}+q-1+C_{2})\ketbra{\nu_{2}}{\nu_{2}},\nonumber\\
 \bar{\tau}_{2}^{\rm opt}&=&(\tau_{11}+q-1+C_{1})\ketbra{\nu_{1}}{\nu_{1}}+\tau_{12}\ketbra{\nu_{1}}{\nu_{2}}+\tau_{21} \ketbra{\nu_{2}}{\nu_{1}}+(\tau_{22}+q-C_{2})\ketbra{\nu_{2}}{\nu_{2}}.~~
\end{eqnarray}
By the complementary slackness condition (iii) of \eqref{eq:optc1}, the every rank of
$\{\bar{M}_{i}^{\rm opt},\bar{\tau}_{i}^{\rm opt}\}_{i=0}^{2}$ is
one. Therefore their determinants become 0, which means
\begin{equation}
|\tau_{12}|=\sqrt{\tau_{11}\tau_{22}}\quad \mbox{and}\quad
\bar{M}_{i}^{\rm opt}=\tr[\bar{M}_{i}^{\rm opt}]\cdot\left[I_{2}-\frac{\bar{\tau}_{i}^{\rm
opt}}{\tr[\bar{\tau}_{i}^{\rm opt}]}\right]\ \forall i.
\end{equation}
Then, we have
$\tau_{11}=\lambda_{1}$ and $\tau_{22}=\lambda_{2}$. Since
$\tr[\bar{M}_{i}^{\rm opt}]$ is the probability that $M_{i}^{\rm
opt}$ may be detected, $\tr[\bar{M}_{0}^{\rm opt}]$ becomes $P_{\rm
I}(q)$. The phase of $\tau_{12}$ and the form of
$\tr[\bar{M}_{i}^{\rm opt}]$ can be obtained by condition (i). The
completeness condition of the POVM is represented as
\begin{equation}
\frac{\tr[\bar{M}_{0}^{\rm opt}]}{\tr[\bar{\tau}_{0}^{\rm opt}]}\cdot\bar{\tau}_{0}^{\rm opt}+
\frac{\tr[\bar{M}_{1}^{\rm opt}]}{\tr[\bar{\tau}_{1}^{\rm opt}]}\cdot\bar{\tau}_{1}^{\rm opt}+
\frac{\tr[\bar{M}_{2}^{\rm opt}]}{\tr[\bar{\tau}_{2}^{\rm opt}]}\cdot\bar{\tau}_{2}^{\rm opt}
=I_{2}-\rho_{0}. 
\end{equation}
$\rho_{12}$ and $\tau_{12}$ have the relation of
$\rho_{12}/\tau_{12}=-\sum_{i=0}^{2}(\tr[\bar{M}_{i}^{\rm
opt}]/\tr[\bar{\tau}_{i}^{\rm opt}])$. By $M_{i}^{\rm
opt}\neq0(\forall i)$ and the non-negativity of POVM, the right hand
side of the equation is always negative, and we get
$\rho_{12}/\tau_{12}=-|\rho_{12}|/|\tau_{12}|$. That is,
$\tau_{12}=-(\rho_{12}/|\rho_{12}|)\sqrt{\lambda_{1}\lambda_{2}}$.
And $\bar{P}_{\rm cor}^{\rm opt}(q)$ is found as \eqref{eq:pcpi2}.
Then we have $\tr[\bar{M}_{i}^{\rm opt}]=\eta_{i}$ by the following
relation:
\begin{eqnarray}
\left(\begin{array}{c}
\tr[\bar{M}_{0}^{\rm
opt}]\\\tr[\bar{M}_{1}^{\rm
opt}]\\\tr[\bar{M}_{2}^{\rm
opt}]\\
\end{array}\right)
=
\left(\begin{array}{ccc}
1&1&1\\
\frac{1}{\tr[\bar{\tau}_{0}^{\rm opt}]} & \frac{1}{\tr[\bar{\tau}_{1}^{\rm opt}]} & \frac{1}{\tr[\bar{\tau}_{2}^{\rm opt}]}\\
\frac{\lambda_{1}}{\tr[\bar{\tau}_{0}^{\rm opt}]} & \frac{\lambda_{1}+q-C_{1}}{\tr[\bar{\tau}_{1}^{\rm opt}]} & \frac{\lambda_{1}+q -1 +C_{1}}{\tr[\bar{\tau}_{2}^{\rm opt}]}
\end{array}\right)^{-1}
\left(\begin{array}{c}
1\\ \frac{|\rho_{12}|}{\sqrt{\lambda_{1}\lambda_{2}}}\\ \rho_{22}\\
\end{array}\right).
\end{eqnarray}
 Therefore $\bar{M}_{i}^{\rm opt}$ is represented as \eqref{eq:optm2}.
The result implies the following. If $\lambda_{i}\geq0(\forall i)$
and $\eta_{i}>0(\forall i)$, we have $M_{i}^{\rm opt}\neq 0(\forall
i)$. Otherwise,
we find $M_{1}^{\rm opt}=0$ or $M_{2}^{\rm opt}=0$. \hfill$\square$
\section{Proof of Uniqueness of Optimal POVM}\label{app:ano}
Here we prove the following fact: When $\{q_{i}{\bm
v}_{i}\}_{i=0}^{2}$ forms a triangle, if $r_{i}^{\rm opt}\neq 0
(\forall i)$, then the POVM $\{M_{i}=p_{i}(I_2+{\bm
u}_{i}\cdot{\bm\sigma})\}_{i=0}^{2}$ fulfilling the optimality
condition \eqref{eq:optc2} is unique. For the proof, we use ${\bm
v}_{0}$ as an arbitrary Bloch vector extrinsic to ${\bm v}_{1}$,${\bm v}_{2}$.
Since $M_{k}^{\rm opt}=I_2$ implies $r_{k}^{\rm opt}=0$, at least two of $\{M_{i}^{\rm opt}\}_{i=0}^{2}$ are nonzero. \\
\indent First, we consider the case that there exists
$\{p_{i}\neq0,{\bm u}_{i}\}_{i=0}^{2}$ and $\{r_{i}\neq0,{\bm
w}_{i}\}_{i=0}^{2}$ fulfilling optimality condition
\eqref{eq:optc2}. Without loss of generality, we can set $q_{0}\geq
q_{1},q_{2}$. Then (iii) becomes ${\bm u}_{i}\cdot{\bm
w}_{i}=-1(\forall i)$. This can be rewritten as $\|{\bm
u}_{i}\|_{2}=1$, ${\bm w}_{i}=-{\bm u}_{i}(\forall i)$, and (ii) is
as follows: $r_{i}-r_{0}=e_{i}$, ${\bf R}\equiv q_{i}{\bm
v}_{i}-r_{i}{\bm u}_{i}$$(i=0,1,2)$. $e_{i}$ is the difference
between two prior probabilities $q_{0}$ and $q_{i}$. This condition
means the following; $\{r_{i}{\bm u}_{i}\}_{i=0}^{2}$ forms a
triangle congruent to a triangle $\{q_{i}{\bm v}_{i}\}_{i=0}^{2}$,
and $\{r_{i}{\bm u}_{i}\}_{i=0}^{2}$ coincides with $\{q_{i}{\bm
v}_{i}\}_{i=0}^{2}$ by parallel transport ${\bf R}$. Then (i)
contains the following statement. ${\bf R}$ lies in the interior of
the triangle $\{q_{i}{\bm v}_{i}\}_{i=0}^{2}$, and the distance from
this point to the vertex of the triangle $q_{i}{\bm v}_{i}$ is
$r_{i}$. The points fulfilling $r_{i}-r_{0}=e_{i}$ satisfy the
following hyperbolic equation:
\begin{eqnarray}\label{eq:r0}
r_{0}=\frac{l_{i}^{2}-e_{i}^{2}}{2(l_{i}\cos\theta_{i}+e_{i})}.
\end{eqnarray}
Above $l_{i}$ is the distance between two vectors $q_{0}{\bm v}_{0}$
and $q_{i}{\bm v}_{i}$, and $\theta_{i}$ is the angle between two
sides $\{{\bf R},q_{0}{\bm v}_{0}\}$ and $\{q_{0}{\bm
v}_{0},q_{i}{\bm v}_{i}\}$. As $\theta_{i}$ increases, $r_{0}$ also
increases, and inside the triangle $\{q_{i}{\bm v}_{i}\}_{i=0}^{2}$
the position of ${\bf R}$ is unique. This means that the
$\{p_{i},{\bm u}_{i}\}_{i=0}^{2}$ are unique. Therefore, the optimal
POVM in which every element is nonzero is unique. To make a
distinction, we denote this POVM as $\{M_{i}'\}_{i=0}^{2}$. Suppose
that there exists another POVM satifying the optimality condition
and denote it as $\{M_{i}''\}_{i=0}^{2}$. Then the POVM consisting
of $M_{i}=\epsilon M_{i}'+(1-\epsilon)M_{i}''(0<\epsilon<1)$ is
optimal, and we have $M_{i}\neq 0(\forall i)$. This is contradictory, and therefore the optimal POVM is unique. \\
\indent Next, we consider the case that there exist $\{p_{i},{\bm
u}_{i}\}_{i=0}^{2}$ and $\{r_{i}\neq0,{\bm w}_{i}\}_{i=0}^{2}$
fulfilling optimality condition \eqref{eq:optc2} and one of
$\{p_{i}\}_{i=0}^{2}$ is zero and the others are nonzero. Without
loss of generality, we can set $p_{0}=0$. Then (iii) becomes ${\bm
u}_{i}\cdot{\bm w}_{i}=-1(i=1,2)$. This can turn into $\|{\bm
u}_{i}\|_{2}=1$, ${\bm w}_{i}=-{\bm u}_{i}(i=1,2)$, and (ii) can be
expressed in the following way: $r_{1}-r_{2}=q_{2}-q_{1}$, ${\bf
R}\equiv q_{1}{\bm v}_{1}-r_{1}{\bm u}_{1}=q_{2}{\bm
v}_{2}-r_{2}{\bm u}_{2}$. This condition implies that $\{r_{i}{\bm
u}_{i}\}_{i=1}^{2}$ coincides with the line segment $\{q_{i}{\bm
v}_{i}\}_{i=1}^{2}$ by parallel translation ${\bf R}$. (i) means
that ${\bf R}$ lies in the interior of $\{q_{i}{\bm
v}_{i}\}_{i=1}^{2}$ and the distance from the point to $q_{i}{\bm
v}_{i}$ is $r_{i}$. That is, we have $r_{1}+r_{2}=l_{12}$. $l_{12}$
is the distance between two vectors $q_{1}{\bm v}_{1}$ and
$q_{2}{\bm v}_{2}$. Then $r_{1}$ and $r_{2}$ satisfying
$r_{1}-r_{2}=q_{2}-q_{1}$ are apparently unique. This implies that
$\{p_{i},{\bm u}_{i}\}_{i=1}^{2}$ are unique. Therefore
the optimal POVM satisfing $M_{0}=0$,$M_{1}\neq0$,$M_{2}\neq0$ is unique. To
differentiate from the other POVM, we represent this POVM as
$\{M_{i}'\}_{i=0}^{2}$. We assume that there exists a POVM
satisfying $M_{0}\neq0$ and the optimality condition, and denote it
as  $\{M_{i}''\}_{i=0}^{2}$. Then the POVM consisting of
$M_{i}=\epsilon M_{i}'+(1-\epsilon)M_{i}''$($0<\epsilon<1$) is
optimal. The result is that POVM fulfilling $M_{i}\neq 0 (\forall
i)$ and geometric optimality condition is not unique.
This contradicts the previous result, and the optimal POVM is unique.\\
\indent In conclusion, when $\{q_{i}{\bm v}_{i}\}_{i=0}^{2}$ forms a triangle and $r_{i}^{\rm opt}\neq 0 (\forall i)$, the optimal POVM is unique.\hfill$\square$

\end{document}